\RequirePackage{ifpdf}
\ifpdf 
\documentclass[pdftex]{sigma}
\else
\documentclass{sigma}
\fi

\newcommand{\1}{{(1)}}

\newcommand{\vx}{{\vec{x}}}
\newcommand{\vy}{{\vec{y}}}

\newcommand{\vep}{{\varepsilon}}
\newcommand{\ep}{{\epsilon}}

\newcommand{\cl}{{\cal L}}
\newcommand{\cg}{{\cal G}}
\newcommand{\dirac}{{\delta(\vx - \vy)}}

\numberwithin{equation}{section}

\begin{document}

\allowdisplaybreaks

\renewcommand{\thefootnote}{$\star$}

\renewcommand{\PaperNumber}{059}

\FirstPageHeading

\ShortArticleName{Noncommutativity and Duality through the Symplectic Embedding Formalism}

\ArticleName{Noncommutativity and Duality \\ through the Symplectic Embedding Formalism\footnote{This paper is a
contribution to the Special Issue ``Noncommutative Spaces and Fields''. The
full collection is available at
\href{http://www.emis.de/journals/SIGMA/noncommutative.html}{http://www.emis.de/journals/SIGMA/noncommutative.html}}}

\Author{Everton M.C. ABREU~$^\dag$, Albert C.R. MENDES~$^\ddag$ and Wilson OLIVEIRA~$^\ddag$}

\AuthorNameForHeading{E.M.C. Abreu, A.C.R. Mendes and W. Oliveira}

\Address{$^\dag$~Grupo de F\' isica Te\'orica e Matem\'atica F\' isica, Departamento de F\'{\i}sica,\\
\hphantom{$^\dag$}~Universidade Federal Rural do Rio de Janeiro, BR 465-07, 23890-971, Serop\'edica, RJ, Brazil}
\EmailD{\href{mailto:evertonabreu@ufrrj.br}{evertonabreu@ufrrj.br}}

\Address{$^\ddag$~Departamento de F\'{\i}sica, ICE, Universidade Federal de Juiz de Fora,\\
\hphantom{$^\ddag$}~36036-330, Juiz de Fora, MG, Brazil}
\EmailD{\href{mailto:albert@fisica.ufjf.br}{albert@fisica.ufjf.br}, \href{mailto:wilson@fisica.ufjf.br}{wilson@fisica.ufjf.br}}

\ArticleDates{Received March 30, 2010, in f\/inal form July 05, 2010;  Published online July 21, 2010}

\Abstract{This work is devoted to review the gauge embedding of either commutative and noncommutative (NC) theories using the symplectic formalism framework.  To sum up the main features of the method, during the process of embedding, the inf\/initesimal gauge generators of the gauge embedded theory  are easily and directly chosen. Among other advantages, this enables a greater control over the f\/inal Lagrangian and brings some light on the so-called ``arbitrariness problem''.  This alternative embedding formalism also presents a way to obtain a set of dynamically dual equivalent embedded Lagrangian densities which is obtained after a f\/inite number of steps in the iterative symplectic process, oppositely to the result proposed using the BFFT formalism. On the other hand, we will see precisely that the symplectic embedding formalism can be seen as an alternative and an  ef\/f\/icient procedure to the standard introduction of the Moyal product in order to produce in a na\-tu\-ral way a NC theory.
In order to construct a pedagogical explanation of the method to the nonspecialist we exemplify the formalism showing that the massive NC $U(1)$ theory is embedded in a~gauge theory using this alternative systematic
path based on the symplectic framework.
Further, as other applications of the method, we describe exactly how to obtain a~Lagrangian description for the NC version of some systems reproducing well known theories.  Naming some of them, we use the procedure in the Proca model, the irrotational f\/luid model and the noncommutative self-dual model in order to obtain dual equivalent actions for these theories.  To illustrate the process of noncommutativity introduction we use the chiral oscillator and the nondegenerate mechanics.}


\Classification{70S05; 70S10; 81Q65; 81T75}

\renewcommand{\thefootnote}{\arabic{footnote}}
\setcounter{footnote}{0}

\section{Introduction}

\looseness=-1
The race to investigate  NC f\/ield theories  was mainly motivated by the fact that it was realized that NC spaces naturally arise in string theory with a constant background magnetic f\/ield in the presence of D-branes \cite{witten}.  Today, we believe that noncommutativity is the natural path to understand the physics at the Planck scale, namely the Black Holes quantum mechanics and the physics of the beginning of our Universe (in the Big-Bang).  In this way, it is natural to consider noncommutativity as one of the main ingredients of the quantum gravity investigation.
However, this noncommutativity in the context of the string theory with a constant background magnetic f\/ield in the presence of D-branes was elucidated constructing a mechanical system which reproduces the classical dynamics of strings \cite{string}. This motivated the concept that a~quantum f\/ield theory on NC spacetimes can be realized as a low energy of the theory of open strings. Since then \cite{strings}, NC f\/ield theories have been studied extensively in f\/ield theories
\cite{RB,belov,szabo,omer,hp} as well as other issues in physics \cite{other,alexei,gangopadhyay}.  We mention, among others, $C$, $P$, and $T$ invariance~\cite{Sheikh}, axial anomaly \cite{Ardalan}, noncommutative QED~\cite{Dayi}, supersymmetry~\cite{Girotti}, renormalization and the mixing of infrared and ultraviolet divergences~\cite{mrs,Seiberg}, unitarity~\cite{Gomis}, phenomenology~\cite{Arfaei} and gravity~\cite{reviews}.

Also motivated by planar problems, like the Landau problem of a charged particle moving on a plane embedded in a strong magnetic f\/ield, in \cite{banerjee}, R.~Banerjee discussed how NC structures appear in quantum mechanics on a plane providing a useful way of obtaining them.   It was based on NC algebra in quantum mechanics on a plane that was originated from 't Hooft's analysis on dissipation and quantization \cite{thooft}.
It was demonstrated in~\cite{banerjee} that the coordinates or momenta noncommutativity are in fact dual descriptions, corresponding to distinct polarizations chosen conveniently to transform a second order system into a f\/irst order one.  The Hamiltonian obtained in~\cite{banerjee} following 't Hooft reveals a NC algebra.  Noncommutativity in planar physics, in the context of relativistic spinning particle models modeling anyons have appeared in \cite{planar}.
In this review we will see in detail that following the symplectic formalism the noncommutativity can be introduced naturally and it will be demonstrated in mechanical systems.

In \cite{mrs}  the authors analyzed the IR and UV divergences and showed that the noncommutativity constant, through the Moyal product, is associated with IR/UV mixing, where the physics at high energies af\/fects the physics at low energies.  And this does not occurs in commutative physics.  We will review here that it is also possible to perform a non-perturbative approach and, as a consequence, the noncommutativity constant can be introduced naturally into the theory.

Another way to introduce noncommutativity in quantum mechanics is performing the so-called $\alpha$-deformation of the algebra of classical observables \cite{DJEMAI1}.  The discussion rely on the mapping from classical mechanics to quantum mechanics and consequently to NC quantum mechanics and NC classical mechanics.  This is possible since   quantum mechanics is naturally interpreted as a NC (matrix) symplectic geometry \cite{DJEMAI2}.

In the 80's, mathematicians like A.~Connes, studied and developed NC geometry.  His analysis led to an operator algebraic description of NC spacetimes and motivated the investigation of a~Yang--Mills theory on a NC torus.  Connes developed what is known as a NC standard model which is an extension of the standard model to include a modif\/ied form of general relativity.

Recently, Doplicher, Fredenhagen and Roberts \cite{dfr} brings into the NC literature another motivation for the noncommutativity of spacetime.  Their considerations lead them to construct a bridge between Black Hole formation and the NC spacetime.  The motivation introduced by them is based on the assumption that the NC constant can be interpreted as a coordinate of the system.  After that, Amorim \cite{amorim} introduced its canonically conjugated momentum and construct a quantum mechanics and a f\/ield theory in superior dimensions.

The most popular and underlying concept about the obtainment of NC theories is based on the procedure where one can replace the usual product of f\/ields inside the action by the Moyal product, def\/ined as
\begin{gather*}
\phi_1 (x)\star \phi_2 (x) =\exp\left( {i\over 2} \theta^{\mu \nu}
\partial^{x}_{\mu}\partial^{y}_{\nu} \right) \phi_1 (x) \phi_2 (y)\Big|_{x=y},
\end{gather*}
where $\theta^{\mu \nu}$ is a real and antisymmetric constant matrix\footnote{The case where $\theta^{\mu \nu}$ is a variable of the system with an associated canonical momentum (Amorim's approach) can be seen in another recent review \cite{review2} and in the references therein.}. As a consequence, NC theories are highly nonlocal. We also note that the Moyal product of two f\/ields in the action is the same as the usual product, provided we discarded boundary terms. Thus, the noncommutativity af\/fects just the vertices.

One of the most important ingredients of the usual Standard Model is the concept of gauge symmetry used in its mathematical description.  The so-called gauge theories have played a~fundamental role in f\/ield theories since they are related to the fundamental physical interactions in nature.

In general, those theories have gauge symmetries def\/ined by some relations called, in Dirac's language, f\/irst-class constraints \cite{PD}. The quantization of these theories demands a special care because of the presence of  gauge symmetries indicating some superf\/luous degrees of freedom.  Its elimination can be carried out in some convenient moment before or after any quantization process begins.

Considering second-class systems, which presents dif\/f\/iculties during the covariant quantization because the Poisson brackets are replaced by Dirac brackets, the quantization is contaminated by serious problems such as
ordering operator problems \cite{order} and anomalies \cite{RR} in the context of nonlinear constrained systems and chiral gauge theories, respectively. In this scenario, it seems that it is more natural and safer to develop the quantization of second-class systems without invoking Dirac brackets. The noninvariant system has been embedded in an extended phase space in order to change the second-class nature of constraints to f\/irst-class one.

As it is clear from the title of this paper, we will try to explore here, in a self-sustained and pedagogical way, all the possibilities of utilization of the embedding mechanism concerning NC theories.  Namely, in the obtainment of dual theories equivalent to an NC original theory or through the introduction of noncommutativity into a~commutative f\/ield theory, as said above.

Having said that, it seems natural to us to talk about this formalism strongly connected to gauge theories and which was f\/irstly suggested by Faddeev and Shatashivilli~\cite{FS}.  From the beginning it has been a successful constraint conversion procedure over the last decades.

The main concept behind this procedure resides in the enlargement of phase space with the introduction of new variables, called Wess--Zumino (WZ) variables, that changes the second-class constraint's nature to be a f\/irst-class one. This procedure has been explored in dif\/ferent contexts in order to avoid some problems that af\/fect the quantization process of some theories.  For instance, the chiral theory, where the anomaly obstructs the quantization mechanism and nonlinear models, where the operator ordering ambiguities arise \cite{CNWO,IJMP,BFFT,ANO}.  As we said before, the main proposition of the embedding procedure, being it applied to commutative or NC theories, is to unveil the origin of the ambiguities of all embedding approaches.  With this in mind we will now describe one of the targets of the embedding method.

A few years back, in  \cite{RB} the authors succeeded in working out the embedded version of the massive NC $U(1)$ theory that was obtained through the BFFT constraint conversion scheme. It was described as a way to obtain a set of second-class constraints and the Hamiltonian which form an involutive system of dynamical quantities. However, both the constraints and the Hamiltonian were expressed as a series of Moyal commutators among the variables belonging to the WZ extended phase space. In this review we try to explain the embedded version for the massive NC $U(1)$ theory where the embedded Hamiltonian density is not expressed as an expansion in terms of WZ variables but as a f\/inite sum showing pedagogically how to use the symplectic embedding formalism \cite{ANO}.  In \cite{ANO,ROTHE}, for example, the authors have used the symplectic formalism in order to embed second-class systems and properly systematize the symplectic embedding formalism, as performed in  BFFT \cite{BT} and iterative \cite{IJMP} methods for example.

One advantage of this method, which was inspired by \cite{ANO}, comes from its simple and direct way of choosing conveniently the inf\/initesimal gauge generators of the built gauge theory. This gives us freedom to choose the content of the embedded symmetry. This feature makes it possible to obtain a great control over the f\/inal Lagrangian. For example, with the BFFT \cite{BFFT} method, NC and non-Abelian theories are usually embedded into theories with inf\/inite terms in the Hamiltonian and with inf\/initesimal gauge generators that cannot be expressed in closed form \cite{kpr,RB}. This can be avoided with the embedding method, because the inf\/initesimal gauge generators are not deduced from previous unclear choices, but instead, are directly chosen.

Comparing more specif\/ically both BFFT and the embedding formalism we can say that the BFFT method succeeded when applied to a great number of important physical models.  Although the basic concepts are equal, they were implemented following dif\/ferent directions.  The BFFT method have ambiguity problems that naturally arises when the second-class constraint is converted into f\/irst-class one with the introduction of WZ variables.  The embedding method is not af\/fected by this ambiguity problem.  This method is based on Faddeev's suggestion \cite{FS} and is elaborated on a framework to deal with noninvariant models, namely the symplectic formalism.  The details and references can be found in \cite{ANO}, where the practical dif\/ferences between both methods are pinpointed precisely during the calculations present there.

Another possibility of performing a kind of ``general embedding" is that, instead of choosing the gauge generators at the beginning, one can leave some unf\/ixed parameters with the aim of f\/ixing them latter, i.e., when the f\/inal Lagrangian has being achieved. Although one can arrive faster at the f\/inal theory f\/ixing such parameters as soon as possible, this path is more interesting in order to study the initial theory and it is helpful if the desired symmetry is unknown, but one can still be looking for some particular aspects of the Lagrangian. This path to ``general embedding" was employed in each one of the applications.

We have to mention that this approach to embedding is not dependent on any undetermined constraint structure and also works for unconstrained systems.  This is dif\/ferent from all the existent embedding techniques that were used
to convert \cite{BFFT,IJMP,cw}, project \cite{wis} or reorder~\cite{mitra} the existent second-class constraints into a f\/irst-class system. This technique on the other hand only deals with the symplectic structure of the theory so that the embedding structure is independent of any pre-existent constrained structure.

This review is based mainly on some publications on  the subject \cite{nossosartigos}.
Hence, we hope we succeeded in constructing a self-contained review to describe the NC and commutative applications of this formalism that generalizes the quantization by deformation introduced in \cite{DJEMAI1}.  We intend to explore, with a new insight, how the NC geometry can be introduced into a commutative f\/ield theory.  Further, this method describes precisely how to obtain a Lagrangian description for the NC version of the system.   To accomplish this, a systematic way to introduce NC geometry into commutative systems, based on the symplectic approach and on the Moyal product is reviewed.    To demonstrate the approach, we use two well known systems, the chiral oscillator and some nondegenerate classical mechanics.  We computed precisely the NC contributions through this generalized symplectic method and obtain exactly the actions in NC space found in literature.   It is important to notice that it is the f\/irst part of a formalism which actual target is to introduce the NC geometry in constrained and non-constrained systems.

Our paper is organized as follows. In Section~\ref{sec:II}, we present an overview of the symplectic embedding formalism. We note that after a f\/inite number of steps of the iterative symplectic embedding process, we obtain an embedded Hamiltonian density. As a consequence, this Hamiltonian density has a f\/inite number of WZ terms, oppositely to \cite{RB}.
In Section~\ref{sec:III}, we exemplify the formalism analyzing the symplectic quantization of the Proca model, which is a classical example of a theory without gauge symmetry due to a mass term. Using the Dirac nomenclature~\cite{Dirac}, it is classif\/ied as a second-class system. On the other hand, the irrotational f\/luid model, our second example, cannot be classif\/ied in the same way, because it does not possess any constraints. Besides that, its Lagrangian has a potential term ($1/\rho$), which could bring, at f\/irst sight, some dif\/f\/iculties to the method. The third and last model of the section is the NC self-dual model which is a topologically massive second-class theory with a Chern--Simons-like term. More important, the f\/ields do not commute~-- a feature that, just like non-Abelian algebra, causes some trouble to other gauge embedding methods~\cite{kpr,RB}. The NC self-dual model, as the non-Abelian self-dual model, has been inside the scope of many recent papers, its properties and dualities (under some limits) still need further investigations.
In Section~\ref{sec:IV} we described the utilization of the method to introduce the noncommutativity to standard theories.  In Subsection~\ref{subsec:IVA} we explain the embedding formalism adapted to noncommutativity.  In Subsection~\ref{subsec:IVB} we explore two examples, the chiral oscillator and an arbitrary nondegenerate mechanics, as we said just above.   At the end of this last section, we will make some concluding remarks and some perspectives. In the Appendix, we list some properties of the Moyal product that we use in this paper.

\section{The formalism}
\label{sec:II}

In this section, we describe the alternative embedding technique that changes the second-class nature of constrained systems to f\/irst-class one.  As we said in the last section, this technique follows the Faddeev and Shatashivilli idea \cite{FS} and is based on a contemporary framework that handles constrained models, namely, the symplectic formalism \cite{FJ,BC}.

In order to systematize the symplectic embedding procedure, we consider a general noninvariant mechanical model whose dynamics is governed by a Lagrangian ${\cal L}(a_i,\dot a_i,t)$ (with $i=1,2,\dots,N$), where $a_i$ and $\dot a_i$ are space and velocity variables, respectively. Notice that this model does not result in a loss of generality or physical content. Following the symplectic method the zeroth-iterative f\/irst-order Lagrangian 1-form is written as
\begin{equation}
\label{2000}
{\cal L}^{(0)}dt = A^{(0)}_\theta d\xi^{(0)\theta} - V^{(0)}(\xi)dt,
\end{equation}
where the symplectic variables are
\begin{gather*}
\xi^{(0)\alpha} =  \left\{ \begin{array}{ll}
                               a_i, & \mbox{with $\alpha=1,2,\dots,N $,} \\
                               p_i, & \mbox{with $\alpha=N + 1,N + 2,\dots,2N ,$}
                           \end{array}
                     \right.
\end{gather*}
$A^{(0)}_\alpha$ are the canonical momenta and $V^{(0)}$ is the symplectic potential. The symplectic tensor is given by
\begin{gather}
\label{2010}
f^{(0)}_{\alpha\beta} = {\partial A^{(0)}_\beta\over \partial \xi^{(0)\alpha}}
-{\partial A^{(0)}_\alpha\over \partial \xi^{(0)\beta}}.
\end{gather}
When the two-form $f \equiv \frac{1}{2}f_{\theta\beta}d\xi^\theta \wedge d\xi^\beta$ is singular, the symplectic matrix (\ref{2010}) has a zero-mode $(\nu^{(0)})$ that generates a new constraint when contracted with the gradient of the symplectic potential,
\begin{equation*}
\Omega^{(0)} = \nu^{(0)\alpha}\frac{\partial V^{(0)}}{\partial\xi^{(0)\alpha}}.
\end{equation*}
This constraint is introduced into the zeroth-iterative Lagrangian 1-form, (\ref{2000}), through a~Lag\-range multiplier $\eta$, generating the next iteration
\begin{gather*}
{\cal L}^{(1)}dt  =  A^{(0)}_\theta d\xi^{(0)\theta} + d\eta\Omega^{(0)}- V^{(0)}(\xi)dt
 =  A^{(1)}_\gamma d\xi^{(1)\gamma} - V^{(1)}(\xi)dt,
\end{gather*}
with $\gamma=1,2,\dots,(2N + 1)$ and
\begin{gather*}
V^{(1)} = V^{(0)}\big|_{\Omega^{(0)}= 0},\qquad
\xi^{(1)_\gamma}  =  \big(\xi^{(0)\alpha},\eta\big),\qquad
A^{(1)}_\gamma  = \big(A^{(0)}_\alpha, \Omega^{(0)}\big).
\end{gather*}
As a consequence, the f\/irst-iterative symplectic tensor is computed as
\begin{gather*}
f^{(1)}_{\gamma\beta} = {\partial A^{(1)}_\beta\over \partial \xi^{(1)\gamma}}
-{\partial A^{(1)}_\gamma\over \partial \xi^{(1)\beta}},
\end{gather*}
which constitutes a test for the procedure continuation if this tensor is nonsingular, the iterative process stops and the Dirac's brackets among the phase space variables are obtained from the inverse matrix $\big(f^{(1)}_{\gamma\beta}\big)^{-1}$. Consequently, the Hamilton equation of motion can be computed and solved as well~\cite{gotay}.

It is well known that a physical system can be described in terms of a symplectic manifold~$M$, classically at least. From a physical point of view, $M$~is the phase space of the system while a~nondegenerate closed 2-form~$f$ can be identif\/ied as being the Poisson bracket. The dynamics of the system is  determined just specifying a real-valued function (Hamiltonian) $H$ on phase space, i.e., one of these real-valued function solves the Hamilton equation, namely,
\begin{gather}
\label{2050a1}
\iota(X)f = dH,
\end{gather}
and the classical dynamical trajectories of the system in phase space are obtained. It is important to mention that if $f$ is nondegenerate, (\ref{2050a1}) has a unique solution. The nondegeneracy of $f$ means that the linear map $\flat:TM\rightarrow T^*M$ def\/ined by $\flat(X):=\flat(X)f$ is an isomorf\/ism.  Due to this, (\ref{2050a1}) is solved uniquely for any Hamiltonian $(X=\flat^{-1}(dH))$.

On the contrary, the tensor has a zero-mode and a new constraint arises, indicating that the iterative process goes on until the symplectic matrix becomes nonsingular or singular. If this matrix is nonsingular, the Dirac's brackets will be determined. In~\cite{gotay}, the authors consider in detail the case
when $f$ is degenerate, which usually arises when constraints are present in the system. In this case, $(M,f)$ is called presymplectic manifold. As a consequence, the Hamilton equation~(\ref{2050a1}), may or may not possess solutions, or possess nonunique solutions. Oppositely, if this matrix is singular and the respective zero-mode does not generate new constraints, the system has a symmetry.

The systematization of the symplectic embedding formalism begins by assuming that the gauge invariant version of the general Lagrangian $({\tilde {\cal L}}(a_i,\dot a_i,t))$ is given by
\begin{gather}
\label{2051}
{\tilde{\cal  L}}(a_i,\dot a_i,\varphi_p,t) = {\cal L}(a_i,\dot a_i,t) + {\cal L}_{\rm WZ}(a_i,\dot a_i,\varphi_p),\qquad p = 1,2,
\end{gather}
where $\varphi_p = (\theta, \dot\theta)$ and the extra term $({\cal L}_{\rm WZ})$ depends on the original $(a_i,\dot a_i)$ and WZ $(\varphi_p)$ conf\/iguration variables. Indeed, this WZ Lagrangian can be expressed as an expansion in orders of the WZ variable $(\varphi_p)$ such as
\begin{gather*}
{\cal L}_{\rm WZ}(a_i,\dot a_i,\varphi_p) = \sum_{n=1}^\infty \upsilon^{(n)}(a_i,\dot a_i, \varphi_p),\qquad \text{with}\ \  \upsilon^{(n)}(\varphi_p)\sim \varphi_p^{n},
\end{gather*}
which satisf\/ies the following boundary condition,
\begin{gather*}
{\cal L}_{\rm WZ} (\varphi_p=0)= 0.
\end{gather*}

The reduction of the Lagrangian in (\ref{2051}), into f\/irst order form precedes the beginning of the conversion process, thus
\begin{gather}
\label{2052}
{\tilde{\cal L}}^{(0)}dt = A^{(0)}_{\tilde\alpha}d\tilde\xi^{(0){\tilde \alpha}} + \pi_\theta d\theta - {\tilde V}^{(0)}dt,
\end{gather}
where $\pi_\theta$ is the canonical momentum conjugated to the WZ variable, that is,
\begin{gather}
\label{2052aa2}
\pi_\theta = \frac{\partial{\cal L}_{WZ}}{\partial\dot\theta} = \sum_{n=1}^\infty \frac{\partial\upsilon^{(n)}(a_i, \dot a_i,\varphi_p)}{\partial\dot\theta}.
\end{gather}

The expanded symplectic variables are $\tilde\xi^{(0){\tilde \alpha}} \equiv (a_i, p_i,\varphi_p)$ and the new symplectic potential becomes
\begin{gather*}
{\tilde V}^{(0)} = V^{(0)} + G(a_i,p_i,\lambda_p),\qquad p = 1,2,
\end{gather*}
where $\lambda_p=(\theta,\pi_\theta)$. The arbitrary function $G(a_i,p_i,\lambda_p)$ is expressed as an expansion in terms of the WZ f\/ields, namely
\begin{gather*}
G(a_i,p_i,\lambda_p)= \sum_{n=0}^\infty{\cal G}^{(n)}(a_i,p_i,\lambda_p),
\end{gather*}
with
\begin{gather*}
{\cal G}^{(n)}(a_i,p_i,\lambda_p) \sim \lambda_p^n.
\end{gather*}
In this context, the zeroth-iteration canonical momenta are given by
\begin{gather*}
{\tilde A}_{\tilde\alpha}^{(0)} = \left\{\begin{array}{ll}
                                  A_{\alpha}^{(0)}, & \mbox{with} \ \tilde\alpha  =1,2,\dots,N,\\
                                  \pi_\theta, & \mbox{with} \ \ {\tilde\alpha}= N + 1,\\
                                   0, & \mbox{with} \ \ {\tilde\alpha}= N + 2.
                                    \end{array}
                                  \right.
\end{gather*}
The corresponding symplectic tensor, obtained from the following general relation
\begin{gather*}
{\tilde f}_{\tilde\alpha\tilde\beta}^{(0)} = \frac {\partial {\tilde A}_{\tilde\beta}^{(0)}}{\partial \tilde\xi^{(0)\tilde\alpha}} - \frac {\partial {\tilde A}_{\tilde\alpha}^{(0)}}{\partial \tilde\xi^{(0)\tilde\beta}},
\end{gather*}
is
\begin{gather}
\label{2076b}
{\tilde f}_{\tilde\alpha\tilde\beta}^{(0)} = \left( \begin{array}{ccc}
{f}_{\alpha\beta}^{(0)} & 0  & 0 \\
0 & 0 & - 1 \\
0 & 1 & 0
\end{array} \right) ,
\end{gather}
which should be a singular matrix.

The implementation of the symplectic embedding scheme consists in computing the ar\-bit\-rary function $G(a_i,p_i,\lambda_p)$. To accomplish this, the correction terms in orders of $\lambda_p$, inside ${\cal G}^{(n)}(a_i,p_i,\lambda_p)$, must be computed as well. If the symplectic matrix, (\ref{2076b}), is singular, it has a~zero-mode $\tilde\varrho$ and, consequently, we have
\begin{gather}
\label{2076}
\tilde\varrho^{(0)\tilde\alpha}{\tilde f}^{(0)}_{\tilde\alpha\tilde\beta} = 0,
\end{gather}
where we assume that this zero-mode is
\begin{gather}
\label{2076a}
\tilde\varrho^{(0)}=\left(\begin{array}{ccc}\gamma^\alpha & 0 & 0 \end{array}\right),
\end{gather}
where $\gamma^\alpha$, is a generic line matrix. Using the relation given in (\ref{2076}) together with~(\ref{2076b}) and~(\ref{2076a}), we have that
\begin{gather*}
\gamma^\alpha{ f}_{\alpha\beta}^{(0)} = 0.
\end{gather*}

In this way, a zero-mode is obtained and, in agreement with the symplectic formalism, this zero-mode must be multiplied by the gradient of the symplectic potential, namely,
\begin{gather*}
\tilde\varrho^{(0)\tilde\alpha}\frac{\partial \tilde V^{(0)}}{\partial \tilde\xi^{(0)\tilde\alpha}} = 0.
\end{gather*}
As a consequence, a constraint arises as being
\begin{gather*}
\Omega = \gamma^\alpha\left[\frac{\partial V^{(0)}}{\partial \xi^{(0)\alpha}} + \frac{\partial G(a_i,p_i,\lambda_p)}{\partial \xi^{(0)\alpha}}\right].
\end{gather*}
Due to this, the f\/irst-order Lagrangian is rewritten as
\begin{gather*}
{\tilde{\cal L}}^{(1)} = A^{(0)}_{\tilde\alpha}\dot{\tilde\xi}^{(0){\tilde \alpha}} + \pi_\theta\dot\theta + \Omega \dot\eta - {\tilde V}^{(1)},
\end{gather*}
where ${\tilde V}^{(1)} = V^{(0)}$. Note that the symplectic variables are now $\tilde\xi^{(1)\tilde\alpha}\equiv (a_i,p_i,\eta,\lambda_p)$ (with $\tilde\alpha = 1,2,\dots,N+3$) and the corresponding symplectic matrix becomes
\begin{gather}
\label{2078}
{\tilde f}_{\tilde\alpha\tilde\beta}^{(1)} = \left( \begin{array}{cccc}
{ f}_{\alpha\beta}^{(0)} &  {f}_{\alpha\eta} & 0 & 0 \\
{f}_{\eta\beta} & 0 & {f}_{\eta\theta} & {f}_{\eta\pi_\theta} \\
0 & {f}_{\theta\eta} & 0 & -1 \\
0 & {f}_{\pi_\theta\eta} & 1 & 0
\end{array} \right) ,
\end{gather}
where
\begin{gather*}
{f}_{\eta\theta} = -\frac{\partial}{\partial\theta}\left[ \gamma^\alpha\left(\frac{\partial V^{(0)}}{\partial \xi^{(0)\alpha}} + \frac{\partial G(a_i,p_i,\lambda_p)}{\partial \xi^{(0)\alpha}}\right)\right]
,\nonumber\\
{f}_{\eta\pi_\theta} = -\frac{\partial}{\partial{\pi_\theta}}\left[\gamma^\alpha\left( \frac{\partial V^{(0)}}{\partial \xi^{(0)\alpha}} + \frac{\partial G(a_i,p_i,\lambda_p)}{\partial \xi^{(0)\alpha}}\right)\right],\\ 
{f}_{\alpha\eta} = \frac{\partial \Omega}{\partial \xi^{(0)\alpha}} = \frac{\partial}{\partial \xi^{(0)\alpha}}\left[\gamma^\alpha\left(\frac{\partial V^{(0)}}{\partial\xi^{(0)\alpha}} + \frac{\partial G(a_i,p_i,\lambda_p)}{\partial\xi^{(0)\alpha}}\right)\right]
.\nonumber
\end{gather*}

Since our goal is to unveil a WZ symmetry, this symplectic tensor must be singular and, consequently, it has a zero-mode, namely,
\begin{gather}
\label{2078b}
\tilde \nu^{(1)}_{(\nu)(a)} = \left(\begin{array}{cccc}\mu^\alpha_{(\nu)} & 1 & a & b \end{array} \right),
\end{gather}
which satisf\/ies the relation
\begin{gather}
\label{2078c}
{\tilde \nu}^{(1)\tilde\alpha}_{(\nu)(a)}{\tilde f}_{\tilde\alpha\tilde\beta}^{(1)}  = 0.
\end{gather}
Note that the parameters $(a,b)$ can be 0 or 1 and $\nu$ indicates the number of choices for ${\tilde \nu}^{(1)\tilde\alpha}$. It is important to notice that $\nu$ is not a f\/ixed parameter. As a consequence, there are two independent set of zero-modes, given by
\begin{gather}
\tilde \nu^{(1)}_{(\nu)(0)}  =  \left(\begin{array}{cccc}\mu^\alpha_{(\nu)} & 1 & 0 & 1\end{array} \right),\qquad
\tilde \nu^{(1)}_{(\nu)(1)} = \left(\begin{array}{cccc}\mu^\alpha_{(\nu)} & 1 & 1 & 0\end{array} \right).\label{2078d}
\end{gather}
The matrix elements $\mu^\alpha_{(\nu)}$ have some arbitrariness which can be f\/ixed in order to disclose a desired WZ gauge symmetry. In addition, in the formalism the zero-mode $\tilde\nu^{(1)\tilde\alpha}_{(\nu)(a)}$ is the gauge symmetry generator, which allows us to display the symmetry from the geometrical point of view.
At this point, we stress upon the fact that this is an important feature since it opens up the possibility to disclose the desired hidden gauge symmetry from the noninvariant model. Dif\/ferent choices of zero-mode generates dif\/ferent gauge invariant versions of the second-class system.  However, these gauge invariant descriptions are dynamically
tantamount, i.e., there is the possibility to relate this set of independent zero-modes, equation~(\ref{2078d}), through canonical transformation $(\tilde {\bar\nu}^{(\prime,1)}_{(\nu)(a)} = T . \tilde {\bar\nu}^{(1)}_{(\nu)(a)})$ where bar means transpose matrix, for example,
\begin{gather*}
\left(\begin{array}{c}\mu^\alpha_{(\nu)} \\ 1 \\ 0 \\ 1\end{array} \right) =
\left(\begin{array}{cccc}1 & 0 & 0 & 0\\
0 & 1 & 0 & 0\\
0 & 0 & 0 & 1\\
0 & 0 & 1 & 0\end{array} \right)
\left(\begin{array}{c}\mu^\alpha_{(\nu)} \\ 1 \\ 1 \\ 0\end{array} \right).
\end{gather*}
It is important to mention here that, in the context of the BFFT formalism, dif\/ferent choices for the degenerated matrix $X$ lead to dif\/ferent gauge invariant versions of the second-class model~\cite{BN}. Now, it becomes clear that the arbitrariness present in BFFT method and within the iterative constraint conversion methods has its origin on the choice of the zero-mode.  It is important to notice that the choice of a zero-mode is not totally arbitrary since of course there are physical conditions that, at a certain point of the process, permit or forbid certain zero-mode choices.  It can be seen with details in~\cite{podolsky}.

From relation (\ref{2078c}), together with (\ref{2078}) and (\ref{2078b}), some dif\/ferential equations involving  $G(a_i,p_i,\lambda_p)$ are obtained, namely,
\begin{gather*}
0 = \mu^\alpha_{(\nu)}{ f}_{\alpha\beta}^{(0)} + { f}_{\eta\beta},\nonumber\\
0 = \mu^\alpha_{(\nu)}{ f}_{\alpha\eta}^{(0)} + a { f}_{\theta\eta} + b { f}_{\pi_\theta\eta},\nonumber\\
0 = { f}_{\eta\theta}^{(0)} + b, \\ 
0 = { f}_{\eta\pi_\theta}^{(0)} - a.\nonumber
\end{gather*}
Solving the relations above, some correction terms, within $\sum_{m=0}^\infty {\cal G}^{(m)}(a_i,p_i,\lambda_p)$, can be determined, also including the boundary conditions $({\cal G}^{(0)}(a_i, p_i,\lambda_p = 0 ))$.

In order to compute the remaining corrections terms for $G(a_i,p_i,\lambda_p)$, we impose that no more constraints arise from the multiplication of the zero-mode $(\tilde\nu^{(1)\tilde\alpha}_{(\nu)(a)})$ by the gradient of potential ${\tilde V}^{(1)}(a_i,p_i,\lambda_p)$. This condition generates a general dif\/ferential equation, which reads as
\begin{gather}
0 = \tilde\nu^{(1)\tilde\alpha}_{(\nu)(a)}\frac{\partial {\tilde  V}^{(1)}(a_i,p_i,\lambda_p)}{\partial{\tilde\xi}^{(1)\tilde\alpha}}
\nonumber\\
\phantom{0}= \mu^\alpha_{(\nu)} \left[\frac{\partial {V}^{(1)}(a_i,p_i)}{\partial{\xi}^{(1)\alpha}} + \frac{\partial  G(a_i,p_i,\theta,\pi_\theta)}{\partial{\xi}^{(1)\alpha}}\right] + a \frac{\partial G(a_i,p_i,\lambda_p )}{\partial\theta} + b\frac{\partial G(a_i,p_i,\lambda_p)}{\partial\pi_\theta}\nonumber\\
\phantom{0} =  \mu^\alpha_{(\nu)} \left[\frac{\partial {V}^{(1)}(a_i,p_i)}{\partial{\xi}^{(1)\alpha}} + \sum_{m=0}^\infty\frac{\partial {{\cal G}}^{(m)}(a_i,p_i,\lambda_p )}{\partial{\xi}^{(1)\alpha}}\right] + a \sum_{n=0}^\infty\frac{\partial {\cal G}^{(n)}(a_i,p_i,\lambda_p )}{\partial\theta}\nonumber\\
\phantom{0=}+ b \sum_{m=0}^\infty\frac{\partial {{\cal G}}^{(n)}(a_i,p_i,\lambda_p )}{\partial\pi_\theta}.\label{2080}
\end{gather}
The last relation allows us to compute all correction terms in terms of $\lambda_p$, within ${\cal G}^{(n)}(a_i,p_i,\lambda_p)$. This polynomial expansion in terms of $\lambda_p$ is equal to zero, subsequently, all the coef\/f\/icients for each order of these WZ variables must be identically null. In view of this, each correction term in orders of $\lambda_p$ can be determined as well. For a linear correction term, we have
\begin{gather}
\label{2090}
0 = \mu^\alpha_{(\nu)} \left[\frac{\partial {V}^{(0)}(a_i,p_i)}{\partial{\xi}^{(1)\alpha}} + \frac{\partial {{\cal G}}^{(0)}(a_i,p_i)}{\partial{\xi}^{(1)\alpha}}\right] + a \frac{\partial{\cal G}^{(1)}(a_i,p_i,\lambda_p)}{\partial\theta} + b\frac{\partial{{\cal G}}^{(1)}(a_i,p_i,\lambda_p)}{\partial\pi_\theta},
\end{gather}
where the relation $V^{(1)} = V^{(0)}$ was used. For a quadratic correction term, we can write that,
\begin{gather}
\label{2095}
0 = {\mu}^{\alpha}_{(\nu)}\left[\frac{\partial{\cal G}^{(1)}(a_i,p_i,\lambda_p)}{\partial{\xi}^{(0)\alpha}} \right] + a \frac{\partial{\cal G}^{(2)}(a_i,p_i,\lambda_p)}{\partial\theta}  + b \frac{\partial{{\cal G}}^{(2)}(a_i,p_i,\lambda_p)}{\partial\pi_\theta}.
\end{gather}
From these equations, a recursive equation for $n\geq 2$ is proposed so that,
\begin{gather}
\label{2100}
0 = {\mu}^{\alpha}_{(\nu)}\left[\frac{\partial {\cal G}^{(n - 1)}(a_i,p_i,\lambda_p)}{\partial{\xi}^{(0)\alpha}} \right] + a\frac{\partial{\cal G}^{(n)}(a_i,p_i,\lambda_p)}{\partial\theta} + b \frac{\partial{{\cal G}}^{(n)}(a_i,p_i,\lambda_p)}{\partial\pi_\theta},
\end{gather}
which allows us to compute the remaining correction terms as function of $\theta$ and $\pi_\theta$. This iterative process is successively repeated up to (\ref{2080}) when it becomes identically null or when an extra term ${\cal G}^{(n)}(a_i,p_i,\lambda_p)$ can not be computed. Then, the new symplectic potential is written as
\begin{gather*}
{\tilde  V}^{(1)}(a_i,p_i,\lambda_p) = V^{(0)}(a_i,p_i) + G(a_i,p_i,\lambda_p).
\end{gather*}
For the f\/irst case, the new symplectic potential is gauge invariant. For the second case, due to some corrections terms within $G(a_i,p_i,\lambda_p)$ which are not yet determined, this new symplectic potential is not gauge invariant. As a consequence, there are some WZ counter-terms in the new symplectic potential that can be f\/ixed using Hamilton's equation of motion for the WZ variables~$\theta$ and~$\pi_\theta$ together with the canonical momentum relation conjugated to $\theta$, given in~(\ref{2052aa2}). Due to this, the gauge invariant Hamiltonian is obtained explicitly and the zero-mode ${\tilde\nu}^{(1)\tilde\alpha}_{(\nu)(a)}$ is identif\/ied as being the generator of the inf\/initesimal gauge transformation, given by
\begin{gather}
\label{2120}
\delta{\tilde\xi}^{\tilde\alpha}_{(\nu)(a)} = \varepsilon{\tilde\nu}^{(1)\tilde\alpha}_{(\nu)(a)},
\end{gather}
where $\varepsilon$ is an inf\/initesimal parameter.

\section[Dual equivalence of commutative and noncommutative theories]{Dual equivalence of commutative and noncommutative\\ theories}
\label{sec:III}

We will see now some examples of how to use the symplectic embedding formalism.  The f\/irst two treat commutative theories.  We can say that these examples are a kind of warm up in order to tackle the noncommutative models, i.e., the f\/inal two examples.  And in the next section we will see, as explained in the Introduction that this same formalism can be extended in a~convenient way to introduce NC feature inside commutative models.

\subsection{The Proca model}
\label{subsec:IIA}

The reason to choose this model as the f\/irst one resides in the fact that it is a simple one that has its symmetry broken thanks to the mass term.  This f\/irst two applications will be useful to show how to deal with nonlinear-velocity  Lagrangians and how to achieve a St\"uckelberg-like Lagrangian.  Namely, a shift on $A^\mu$ that transforms it into $A^\mu - \partial^\mu \theta$. To this end, we will search for gauge generators of the type
\begin{gather*}
 \delta_\vep A^\mu = \partial^\mu \vep
\qquad \mbox{and}\qquad
 \delta_\vep \theta = \vep .
\end{gather*}
In fact, it will be accomplished by a more general embedding which has the above structure as a special case. The presence of a temporal derivative acting on the inf\/initesimal parameter, as previously explained, will require the use of a Lagrangian of the type $\tilde \cl_{\theta, \gamma}$.

Before dealing with the gauge embedding method, we will introduce the canonical momenta as new independent f\/ields, modifying the theory such that a linear one appears (otherwise it would not be possible to use a symplectic framework). With the metric $g = \mbox{diag} \left( \begin{array}{cccc}+ & - & - & -\end{array}\right)$, the Proca Lagrangian,
\begin{gather*}
	{\cal L}(A_\mu,\partial_\nu A_\mu)  =  -\frac 14 F^{\mu \nu}F_{\mu \nu} + \frac {m^2}2 A^\mu A_\mu,
\end{gather*}
can be written as
\begin{gather}
	{\cal L}(A_\mu,\partial_\nu A_i, \pi_i, \partial_j \pi_i) =  \pi^i \dot A_i + \frac 12 \pi^i \pi_i - \pi^i\partial_i A_0 - \frac 14 F^{ij}F_{ij} + \frac {m^2}2 A^\mu A_\mu,
	\label{proca2}
\end{gather}
where $\mu=0,1,2,3$, $i=1,2,3$ and $F_{\mu \nu} \equiv \partial_\mu A_\nu - \partial_\nu A_\mu$. Through the Euler--Lagrange equations from the last Lagrangian, one can f\/ind that $\pi_i = F_{i 0}$. If $\pi_i$ is replaced by $F_{i0}$ in last Lagrangian, one returns to the f\/irst one.

The next step is to introduce the Wess--Zumino f\/ields $\theta$ and $\gamma$ (just like in \eqref{2052}):
\begin{gather*}
	\tilde {\cal L} = \pi_i \dot A^i + (\Psi + \gamma) \dot \theta - \tilde V,
\end{gather*}
with
\begin{gather*}
	\tilde V \equiv  - \frac 12 \pi^i \pi_i + \pi^i \partial_i A_0 + \frac 14 F^{ij}F_{ij} - \frac {m^2}2 A^\mu A_\mu + G  + \frac k2 \gamma \gamma .
\end{gather*}
The constant $k$ as well as the functions $G$ and $\Psi$ are still unknown but, by def\/inition, $G$ is zero when $\theta$ is zero (the unitary gauge) and both functions rely on $A_\mu$, $\pi_i$, $\theta$ and its spatial derivatives, which ensures that $\dot \theta = k \gamma$. In order to f\/ind the gauge embedded Lagrangian, all our work resides in f\/ixing $k$ and f\/inding the functions $\Psi$ and $G$.

Let $\tilde \xi^\alpha = (A^0, A^i, \pi^i, \theta, \gamma)$ be the symplectic coordinates, then  $\tilde a_\alpha = (0, \pi_j, 0, \Psi + \gamma, 0)$ are the symplectic momenta (see \eqref{2000}) and
\begin{gather*}
	\tilde f = \left(\begin{array}{ccccc} 0 & 0 & 0 & \frac{\delta \Psi(\vy)}{\delta A^0(\vx)} & 0 \\
	0 & 0 & -g_{ji} \dirac & \frac{\delta \Psi(\vy)}{\delta A^i(\vx)} & 0 \\
	0 & g_{ij} \dirac & 0 & \frac{\delta \Psi(\vy)}{\delta \pi^i(\vx)} & 0 \\
	- \frac{\delta \Psi(\vx)}{\delta A^0(\vy)} & - \frac{\delta \Psi(\vx)}{\delta A^j(\vy)} & - \frac{\delta \Psi(\vx)}{\delta \pi^j(\vy)} & \Theta_{xy} & - \dirac \\
	0 & 0 & 0 & \dirac & 0 \end{array} \right)
\end{gather*}
is the symplectic matrix, whose components are \cite{FJ}
\begin{gather*}
	\tilde f_{\alpha \beta} (\vx,\vy) \equiv \frac{\delta \tilde a_\beta(\vy)}{\delta \tilde \xi^{\alpha}(\vx)} - \frac{\delta \tilde a_\alpha(\vx)}{\delta \tilde \xi^{\beta}(\vy)} ,
\end{gather*}
where
\begin{gather*}
\Theta_{xy} = \frac{\delta \Psi(\vy)}{\delta \theta (\vx)}  -  \frac{\delta \Psi(\vx)}{\delta \theta (\vy)} .
\end{gather*}
Note that $\tilde f$ is a $9 \times 9$ matrix with two spatial indexes in each entry. There is also an implicit time dependence, which comes from the coordinates and momenta. In the above representation of $\tilde f$, some zeros in it are actually null columns, null lines or null matrices.

In the symplectic framework, a theory has gauge symmetry if, and only if, the symplectic matrix is degenerate and its zero-modes does not produce new constraints \cite{BC, mon}. In that case, the components of the zero-modes will be the inf\/initesimal gauge generators. Although $\Psi$ is still an arbitrary function, the presence of Dirac deltas in both  last column and last line severely restrain our possibilities of choosing zero-modes (as explained in the last section). With the purpose of avoiding such restraint condition concerning our choices, before trying to gauge embed the theory, we will insert a constraint into the Lagrangian.

In order to generate a suitable new constraint, let us demand that
\begin{gather*}
	\tilde \nu  = \left(\begin{array}{ccccc}1 & 0_{1 \times 3} & 0_{1 \times 3} & 0 & b\end{array} \right)
\end{gather*}
be the zero-mode of $\tilde f$, where $b$ is a constant. The Proca model (\ref{proca2}), whose symplectic matrix is the one above without the last two lines and columns, has the zero-mode $\nu = \left(\begin{array}{ccc}1 & 0_{1\times 3} & 0_{1 \times 3}\end{array} \right)$, hence $\tilde \nu$ complies with $\tilde \nu = \left(\begin{array}{ccc} \nu & 0 & b\end{array} \right)$.

	The constraint generated by $\nu$ in the Proca model is $\Omega = -\partial_i\pi^i - m^2 A_0$. As we will see, $\tilde \nu$~will produce a constraint which is equal to $\Omega$ when $\gamma = \theta = 0$.

	Demanding $\tilde \nu$ to be a zero-mode of $\tilde f$, one condition for $\Psi$ is found, which is
\begin{gather}
	\frac {\delta \Psi(\vy)}{\delta A^0(\vx)} = -b \dirac .
	\label{psi_a0proca}
\end{gather}

As well known from the symplectic theory, the constraint appears from the following contraction,
\begin{gather}
	\tilde \Omega (\vx) = \int d^3y \; \tilde \nu^\alpha(\vx) \frac{\delta \tilde V(\vy)}{\delta \tilde \xi^\alpha (\vx)} = -\partial_i \pi^i - m^2A_0 + \int d^3y \; \frac{\delta G(\vy)}{\delta A_0 (\vx)}  + bk \gamma,
	\label{modconstproca}
\end{gather}
or, for short, $\tilde \Omega = \Omega + G_0 + bk\gamma$, where $G_0$ is implicitly def\/ined.

Following the standard procedure for handling constraints with the symplectic frame\-work~\cite{BC}, we add $\dot \lambda \tilde \Omega$ to $\tilde \cl$ and treat $\lambda$ as a new independent f\/ield, i.e., a Lagrange multiplier, hence,
\begin{gather*}
	\tilde \cl^{(1)} = \pi^i \dot A_i + (\Psi + \gamma) \dot \theta + \dot \lambda \tilde \Omega - \tilde V.
\end{gather*}
The presence of the constraint in the Lagrangian kinetic term allows us to remove it from the potential term. Nevertheless, this common procedure would be of no use here, therefore no change was accomplished in the potential at all.

Fixing $\tilde \xi^{(1)\alpha} = (A^0, A^i, \pi^i, \theta, \gamma, \lambda)$ as the new symplectic coordinates, where hereafter $\alpha = 1,2,\dots ,10$, and with the help of equation (\ref{psi_a0proca}), the following symplectic matrix can be written as,
\begin{gather*}
\tilde f^{(1)} =\! \left(\begin{array}{@{\,\,}c@{\,\,}c@{\,\,}c@{\,\,}c@{\,\,}c@{\,\,}c@{\,\,}} 0 & 0 & 0 & -b\delta^{(3)} & 0 & \frac {\delta G_0(\vy)}{\delta A_0(\vx)} - m^2\delta^{(3)} \vspace{1mm} \\
	0 & 0 & -g_{ji} \delta^{(3)} & \frac{\delta \Psi(\vy)}{\delta A^i(\vx)} & 0 & \frac {\delta G_0(\vy)}{\delta A^i(\vx)} \vspace{1mm} \\
	0 & g_{ij} \delta^{(3)} & 0 & \frac{\delta \Psi(\vy)}{\delta \pi^i(\vx)} & 0 & \frac {\delta G_0(\vy)}{\delta \pi^i(\vx)} - {\partial}_i^y \delta^{(3)} \vspace{1mm} \\
	b\delta^{(3)} & - \frac{\delta \Psi(\vx)}{\delta A^j(\vy)} & - \frac{\delta \Psi(\vx)}{\delta \pi^j(\vy)} & \Theta_{xy} & - \delta^{(3)} & \frac {\delta G_0(\vy)}{\delta \theta (\vx)} \vspace{1mm} \\
	0 & 0 & 0 & \delta^{(3)} & 0 & bk \delta^{(3)} \vspace{1mm} \\
	- \frac {\delta G_0(\vx)}{\delta A_0(\vy)} + m^2 \delta^{(3)} & \frac {\delta G_0(\vx)}{\delta A^j(\vy)} & {\partial}_j^x \delta^{(3)} - \frac {\delta G_0(\vx)}{\delta \pi^j(\vy)} & - \frac {\delta G_0(\vx)}{\delta \theta(\vy)} & -bk \delta^{(3)} & 0\end{array} \right).
\end{gather*}
For the sake of clarity and compact notation it was convenient to use the notation $\delta^{(3)}$ instead of $\dirac$.

Now we are in position to choose the symmetry that the embedded theory will have.
In accordance with \eqref{2078d},  we can select two independent zero-modes to become the inf\/initesimal gauge generators, which are
\begin{gather}
	\tilde \nu_{(\theta)}   =   \left(\begin{array}{cccccc}a_0 & a \partial^i & c \partial^i & -kb & 0 & 1 \end{array} \right), \nonumber \\
	\tilde \nu_{(\gamma)}   =   \left(\begin{array}{cccccc}1 & 0_{1 \times 3} & 0_{1 \times 3} & 0  & b  & 0\end{array}\right) = \left(\begin{array}{cc} \tilde \nu & 0 \end{array}\right).
	\label{zmproca}
\end{gather}
The values of the constants $a_0$, $a$ and $c$ can be freely selected.  We have to remember that dif\/ferent choices directly correspond to dif\/ferent gauge generators. As it will be shown, the value of $b$ is also free.

We know that other structures of zero-modes are possible, some of which entail correspondence to both WZ f\/ields in each set and, instead of being just constants or spatial derivatives, dependence on $A^\mu$ or $\pi^i$, for example, is also possible.

Although at this stage we could f\/ix the above mentioned constants, selecting some of them to be zero,   simplifying considerably the ef\/fort, we will deal with the problem in the present general form, disclosing a wider symmetry.

Just one condition for $\tilde \nu_{(\gamma)}$ is necessary to assure its zero-mode characteristic, namely,
\begin{gather}
	\frac{\delta G_0 (\vy)}{\delta A_0 (\vx)} = (m^2-b^2k)\dirac.
	\label{G0proca}
\end{gather}

This zero-mode need to be a generator of gauge symmetry, therefore no new constraint may arise from its multiplication with the gradient of the potential.  From equations (\ref{modconstproca}) and (\ref{zmproca}), we see that this condition was automatically fulf\/illed.

Now we will turn our attention to $\tilde \nu_{(\theta)}$. There is a set of nontrivial equations that need to be satisf\/ied in order to $\tilde \nu_{(\theta)}$ to be a zero-mode of $\tilde f^{(1)}$. Instead of evaluating them now, it seems to be easier f\/irstly demand that $\tilde \nu_{(\theta)}$ may not give rise to a new constraint. Hence,
\begin{gather}
	0   =   \int d^3y \; \tilde \nu^\alpha_{(\theta)} (\vx) \; \frac{\delta \tilde V (\vy)}{\delta \tilde \xi^{(1) \alpha}(\vx)} \nonumber \\
	\nonumber
\phantom{0} =  \int d^3y \bigg \{ a_0 \dirac ( -\partial_i \pi^i - m^2A_0 )  + a\partial^i_x \dirac ( \partial^j F_{ij} - m^2 A_i )         \\
 \phantom{0   =}{}
 +   c\partial^i_x  \dirac ( -\pi_i + \partial_i A_0 ) + \rho^\mu_x \frac{\delta G(\vy)}{\delta A^\mu(\vx)} + c\partial^i_x \frac{\delta G(\vy)}{\delta \pi^i(\vx)} - kb \frac{\delta G(\vy)}{\delta \theta(\vx)} \bigg\}. \label{Gproca}
\end{gather}
The index $x$ in $\partial^i$ means that the derivative must be evaluated with respect to $x$ (i.e., $\partial_x^i \equiv \partial / \partial x_i$), and
\begin{gather*}
	\rho^\mu_x \equiv \big( a_0, a\partial^i_x\big).
\end{gather*}

	Equation (\ref{Gproca}) can be solved by considering $G$ as a power series of $\theta$ (and its spatial derivatives). Let $\cg_n$ be proportional to $\theta^n$, so $G = \sum_n \cg_n$. The condition $G(\theta = 0) = 0$ leads to $n \ge 1$. Hence,
\begin{gather*}
	\cg_1 = \frac{\theta}{kb} \big( -a_0 \partial_i\pi^i -m^2\rho^\mu A_\mu - c\partial^i\pi_i + c\partial^i\partial_i A_0\big).
\end{gather*}

The terms
\begin{gather*}
\frac{\delta G(\vy)}{\delta A^\mu(\vx)}\qquad \mbox{and} \qquad \frac{\delta G(\vy)}{\delta \pi^i(\vx)}
\end{gather*}
do not contribute to the computation of $\cg_1$.   However, they contribute to others $\cg_n$'s,  besides
$
\frac{\delta G(\vy)}{\delta \theta(\vx)}$,
which encloses the $\theta$ f\/ield. After some straightforward calculations, one can f\/ind $\cg_2$ (without surface terms) as
\begin{gather*}
	\cg_2 = - \frac 1{2(kb)^2} \big\{ c(2 a_0 + c) \partial_i \theta \partial^i \theta  + m^2 \rho^\mu \theta \rho_\mu \theta \big\}.
\end{gather*}

The absence of $A^\mu$ and $\pi^i$ in $\cg_2$ implies $\cg_n = 0$ for all $n \ge 3$. Thus the function $G$ is completely known, and we can write down the expression for $G_0$, which is,
\begin{gather}
	G_0(\vx) \equiv \int d^3y   \frac {\delta G(\vy)}{\delta A^0 (\vx)} = \frac 1 {kb} \big(c\partial^i \partial_i \theta - m^2 a_0 \theta\big).
	\label{G0proca2}
\end{gather}
Applying this result in (\ref{G0proca}), we have that,
\begin{gather}
	k = \frac {m^2}{b^2} ,
	\label{kproca}
\end{gather}
which f\/ixes $k$ with relation to $b$.

	Our next and f\/inal step in order to construct the gauge embedded Lagrangian is to f\/ind $\Psi$. This can be carried out by demanding that $\tilde \nu_{(\theta)}$ be a zero-mode of $\tilde f^{(1)}$. Among some redundant or trivial equations, there are the following important ones (using (\ref{G0proca2}) and (\ref{kproca})),
\begin{gather}
	c\partial_j^x \dirac + \frac{m^2}{b} \frac{\delta \Psi(\vx)}{\delta A^j(\vy)} = 0,
	\label{psiaproca}
\\
	- a\partial_j^x \dirac + \frac {m^2}b \frac {\delta \Psi (\vx)}{\delta \pi^j(\vy)} + \partial_j^x \dirac = 0,
	\label{psipiproca}
\\
	-b a_0 \dirac + a\partial_x^i \frac{\delta \Psi(\vy)}{\delta A^i(\vx)} + c \partial^i_x \frac {\delta \Psi (\vy)}{\delta \pi^i(\vx)} - \frac {m^2}b \Theta_{xy} - \frac{\delta G_0 (\vx)}{\delta \theta (\vy)} = 0.
	\label{psithetaproca}
\end{gather}
	With (\ref{psi_a0proca}) and (\ref{psiaproca})--(\ref{psithetaproca}), up to an additional function just of $\theta$ (action surface term), $\Psi$ can be determined. The answer is
\begin{gather*}
	\Psi = - \frac b {m^2}  \big\{ m^2 A_0 + c \partial_i A^i + (1 - a)\partial^i \pi_i \big\} .
\end{gather*}

Hence, the gauge version of the Lagrangian for the Proca model was found. Nevertheless, it is more interesting to express it without the momenta $\pi_i$. At f\/irst, note that we can drop the term $\dot \lambda \tilde \Omega$ from $\tilde \cl^{(1)}$ without changing the dynamics (one can always turn on the symplectic algorithm again and f\/ind the constraint $\tilde \Omega$), this will lead us back to $\tilde \cl$. By varying $\tilde \cl$ with respect to $\pi_i$ and using the  Euler--Lagrange equations we f\/ind
\begin{gather*}
	\pi_i = \partial_i A_0 - \dot A_i
+ \frac b {m^2} \big[ (1-a) \partial_i \dot \theta + (a_0 + c) \partial_i \theta  \big] .
\end{gather*}

Note that the momenta are not the original ones (which are $F_{i0}$), but when $\theta$ is removed they are recovered.

Also from the Euler--Lagrange equations, we have
\begin{gather*}
	\gamma = \frac {b^2}{m^2} \dot \theta.
\end{gather*}

Thus, eliminating $\pi_i$ and $\gamma$, the Lagrangian $\tilde \cl$ can be written as
\begin{gather}
	\tilde \cl   =   -  \frac 14 F_{\mu \nu} F^{\mu \nu} + \frac {m^2}2 A^\mu A_\mu \nonumber \\
 \phantom{\tilde \cl   =}{} +  \frac b {m^2} \left\{ -m^2 A_0 \dot \theta + (1-a) \partial_i \dot \theta (\partial^i A_0 - \dot A^i) + a_0 \theta (\partial^i \partial_i A_0 - \partial_i \dot A^i ) + \theta m^2 \rho^\mu A_\mu \right\} \label{procaembedL}   \\
\phantom{\tilde \cl   =}{}		  +  \frac {b^2}{m^4} \left \{ \frac 32 (1\!-\!a)^2 \partial_i \dot \theta \partial^i \dot \theta - \frac 12 a_0^2 \partial_i \theta \partial^i \theta + (1\!-\!a) (a_0 \!+ \!c) \partial_i \dot \theta \partial^i \theta + \frac{m^2}2 \rho^\mu \theta \rho_\mu \theta \right \} + \frac {b^2}{2m^2}\dot \theta \dot \theta . \nonumber
\end{gather}

	From the components of $\tilde \nu_\theta$ and $\tilde \nu_\gamma$ the inf\/initesimal gauge generators of the theory are obtained as,
\begin{gather*}
	\delta_\vep A_0   =   \vep a_0 -  \dot \vep, \qquad
	\delta_\vep A^i   =   - a \partial^i \vep,  \qquad
	\delta_\vep \theta   =   - \frac {m^2}b \vep.
\end{gather*}

	The symplectic formalism assures us that $\tilde \cl$ is invariant under the above transformations for any constants $b$, $a_0$ and $a$ (assuming they have proper dimensions, which are squared mass, mass and unit respectively).

	Usually, terms with more than two derivatives in the Lagrangian are not convenient.
Let us avoid these by f\/ixing $a = 1$.

	To obtain an explicit Lorentz invariance, the constants need to be f\/ixed as $b= m^2$, $a = 1$ and $a_0 = 0$ (alternatively, $b$ could also be $-m^2$). With these values, the Lagrangian turn out to have a St\"uckelberg aspect, that is
\begin{gather*}
	\tilde \cl  =  - \frac 14 F_{\mu \nu} F^{\mu \nu} + \frac {m^2}2 A^\mu A_\mu - m^2A^\mu \partial_\mu \theta + \frac {m^2}2 \partial^\mu \theta \partial_\mu \theta .
\end{gather*}

This result could also be calculated from the analysis of the gauge generators and comparing them to (2.38),
and also with the knowledge that the convenient generators are $\delta_\vep A^\mu = - \partial^\mu \vep $ and $ \delta_\vep \theta = -\vep$ (in order to have that $\delta_\vep(A^\mu + \partial^\mu \theta)=0$).
After all, one could f\/ix the constants as soon as they have appeared, achieving the above results more quickly.

	Concerning the Lagrangian (\ref{procaembedL}), it is not the most general one that can be written with the symplectic embedding method. Others structures of the zero-modes $\tilde \nu_{(\theta)}$ and $\tilde \nu_{(\gamma)}$ are also possible, and their components, together with the components of $\tilde \nu$, could also be f\/ield dependent.

\subsection{Irrotational f\/luid model}
\label{subsec:IIB}

	In this section we will apply the formalism to an unconstrained theory. To carry out this it is necessary to use a Lagrangian of the type $\tilde \cl(\theta)$~\eqref{2051}.

	The irrotational f\/luid model has its dynamics governed by the following Lagrangian density
\begin{equation*}
{\cal L} = - \rho\dot\eta + \frac 12 \rho(\partial_a\eta)(\partial^a\eta) - \frac g\rho,
\end{equation*}
where $a=1,2,\dots,d$ (runs through spatial indexes only), $\rho$ is the mass density, $\eta$ is the velocity potential and $g$ is a constant. Here the metric is Euclidean. This model does not possess neither gauge symmetry nor, constraints in the symplectic sense\footnote{The Dirac algorithm always select the constraints in any linear Lagrangian but, for each constraint, it adds a new f\/ield: the canonical momenta. This roundabout procedure is not present in the symplectic algorithm. See \cite{FJ,mon} for details.}. The above Lagrangian is already linear in the velocity, hence we can proceed directly to the embedding process.

In accordance with the last comments, we will not use the $\gamma$ f\/ield. Hence, the gauge embedded Lagrangian is
\begin{gather*}
{\tilde{\cal L}} = - \rho\dot\eta + \Psi \dot \theta + \frac 12 \rho   \partial_a\eta\partial^a\eta - \frac g\rho - G,
\end{gather*}
where $\Psi\equiv \Psi(\rho, \eta, \theta)$ and $G \equiv G(\rho, \eta, \theta)$.

Setting the symplectic coordinate vector as $\tilde \xi^{ \alpha} = (\rho, \eta, \theta)$, the symplectic momenta and matrix are
\begin{gather*}
\tilde a_\alpha = (0, -\rho, \Psi)
\end{gather*}
 and
\begin{gather*}
{\tilde f }= \left(\begin{array}{ccc}0 & - \delta(\vec x - \vec y) & \frac{\delta\Psi(\vec y)}{\delta\rho(\vec x)} \vspace{1mm}\\
\delta(\vec x - \vec y) & 0 & \frac{\delta\Psi(\vec y)}{\delta\eta(\vec x)} \vspace{1mm} \\
- \frac{\delta\Psi(\vec x)}{\delta\rho(\vec y)} & -  \frac{\delta\Psi(\vec x)}{\delta\eta(\vec y)} & \Theta_{xy}  \end{array}\right).
\end{gather*}
Notice that $\vx$ and $\vy$ are $d$-dimensional vectors and like the previous application we have that,
\begin{gather*}
\Theta_{xy} \equiv \frac{\delta \Psi (\vy)}{\delta \theta (\vx)}  -  \frac{\delta \Psi (\vx)}{\delta \theta (\vy)} .
\end{gather*}

The symplectic method af\/f\/irm that if the symplectic matrix is degenerated and one of its (linearly independent) zero-modes does not produce any new constraints, then the theory has gauge symmetry and the inf\/initesimal gauge generators are given by the components of that zero-mode. Due to the absence of the $\gamma$ f\/ield, there is no ``modif\/ied constraint'' to insert.  So we can go forward to the selection of the zero-mode related to the inf\/initesimal gauge generators. The most general constant zero-mode of $\tilde f^{(0)}$ has the form
\begin{gather*}
	\tilde \nu = \left(\begin{array}{ccc}a & b & 1\end{array}\right).
\end{gather*}
This one imposes the following conditions on $\Psi$,
\begin{gather}
	\frac {\delta \Psi (\vx)}{\delta \rho(\vy)}   =    b \dirac,\qquad
	\frac {\delta \Psi (\vx)}{\delta \eta(\vy)}   =   - a \dirac,\qquad
	\Theta_{xy}   =   0 .\label{fluidpsieqs}
\end{gather}

	If one is not interested in a gauge symmetry related to $\rho$ (i.e., $\delta_\vep \rho = 0$), for example, $a$ could be put equal to zero at this point, simplifying future calculations.

From the equations (\ref{fluidpsieqs}) it is possible to f\/ind $\Psi$ as
\begin{gather*}
	\Psi = b \rho - a \eta + f(\theta),
\end{gather*}
where $f(\theta)$ is an arbitrary function of $\theta$ alone. This function, as one can easily check, only contributes to a surface term for the action, therefore it will not be written anymore.

The last step of the formalism concerning this theory is the calculation of $G$. This function can be found by demanding that $\tilde \nu$ does not give rise to any constraint, that is,
\begin{gather*}
\int d^d y\, \tilde \nu^\alpha(\vec x) \frac{\delta \tilde V(\vec y)}{\delta \tilde \xi^\alpha(\vec x)} = 0,
\end{gather*}
with $\tilde V$ being the potential part of $\tilde {\cal L}$, namely,
\begin{gather*}
	\tilde V = - \frac 12 \rho   \partial_a\eta\partial^a\eta + \frac g\rho + G.
\end{gather*}
Hence
\begin{gather*}
\int d^d y \left\{ a \left ( - \frac 12\partial_a\eta   \partial^a\eta   \dirac  -  \frac g {\rho^2} \dirac + \frac {\delta G (\vy)}{\delta \rho (\vx)} \right ) \right. \nonumber \\
\left. \qquad{} +  b \left ( - \rho   \partial_a \eta   \partial^a \dirac + \frac {\delta G (\vy)}{\delta \eta (\vx)} \right ) + \frac {\delta G (\vy)}{\delta \theta (\vx)} \right\}=0.
\end{gather*}
In this equation, every implicit dependence on space refers to the vector $\vy$.

Expanding $G$ in powers of $\theta$, $G = \sum \cg_n$ with $\cg_n \propto \theta^n$ and $n \ge 1$ (due to $G(\theta=0)=0$), we have that
\begin{gather*}
	\cg_1  =  a \left ( \frac 12 \partial_a\eta   \partial^a\eta   \theta + \frac g {\rho^2} \theta \right ) + b \rho \partial_a \eta   \partial^a \theta, \nonumber\\
	\cg_2  =   - a \left ( -a \frac g {\rho^3} \theta^2 + b \partial^a \eta   \partial_a \theta   \theta \right ) - \frac {b^2 }2 \rho \partial_a \theta   \partial^a \theta, \nonumber\\[0.1in]
	\cg_3   =   a \left ( a^2 \frac g {\rho^4} \theta^3 + \frac {b^2} 2\theta \partial^a \theta \; \partial_a \theta \right ), \qquad
	\cg_n   =   a^n \frac {g}{\rho^{n+1}} \theta^n \qquad \forall \;  n \ge 4.  
\end{gather*}

As $\rho > a \theta$ the series $\sum \cg_n$ converges and we f\/ind the following Lagrangian,
\begin{gather*}
	\tilde \cl = - \rho \dot \eta + (b \rho - a \eta) \dot \theta +  (\rho - a\theta) \left ( \frac 12 \partial_a \eta   \partial^a \eta - b \partial^a \eta   \partial_a \theta + \frac {b^2}2  \partial^a \theta   \partial_a \theta \right ) - \frac g {\rho - a \theta}.
\end{gather*}

The above Lagrangian is invariant under gauge transformations that result from \eqref{2120}
\begin{gather*}
\delta_\vep \rho  =  a \vep,\qquad
\delta_\vep \eta  =  b \vep,\qquad
\delta_\vep \theta =  \vep.  
\end{gather*}

	One can easily check that $\delta_\vep \tilde \cl = 0$ (for $\delta_\vep$ acts like a derivative operator).

\subsection{Noncommutative self-dual model}
\label{subsec:IIC}

Now we will analyze the extension of the symplectic embedding technique into the NC scenario.
With some results from the symplectic formalism we construct a dual theory, with gauge symmetry, equivalent to
the NC self-dual model in $2+1$ dimensions. We use the easiness of the symplectic formalism to deal with inf\/initesimal
gauge generators in order to obtain some generalization for the f\/inal
Lagrangian, which has, as a special case, a St\"uckelberg aspect.
The duality is established without the use of the Seiberg--Witten
map and with no restriction on the powers of the parameter of the
Moyal product.

The f\/inal result is a Lagrangian with the gauge symmetry and the
same ``physics" of the NC self-dual model, without
using any kind of approximation or restriction concerning the Moyal
product. Deliberately some of its parameters are left unf\/ixed.
The reason is that one can analyze the gauge generators of this
Lagrangian, compare these with the desired ones and f\/ix the
parameters accordingly. In the future, these
parameters will be f\/ixed with the aim of obtaining a Lagrangian with a
St\"uckelberg form.

To achieve our objective, what we need is a modif\/ied NC self-dual Lagrangian whose symplectic matrix is degenerated and its zero-modes do not produce new constraints. Nevertheless, this new Lagrangian, with some gauge f\/ixing conditions (the unitary gauge), must be equal to the original one (except for surface terms).

The NC self-dual Lagrangian is
\begin{gather*}
	{\cal L} = \frac 12 f^{\mu} f_{\mu} - \frac 1{4m} \epsilon^{\mu \nu \lambda} f_\mu F_{\nu \lambda},
\end{gather*}
with summation convention implied, $\mu, \nu, \lambda = 0,1,2$, metric
$g= \mbox{diag} \left(\begin{array}{ccc}+ & - & -\end{array}\right)$ and $\epsilon^{012} = 1$. The NC f\/ield nature of this Lagrangian resides solely in
\begin{gather*}
	F_{\mu \nu} = \partial_\mu f_\nu - \partial_\nu f_\mu - ie[f_\mu,f_\nu],
\end{gather*}
where $[\,,\,]$ is the Moyal commutator, i.e.,
\begin{gather*}
	[f_\mu, f_\nu](\vx) = (f_\mu \star f_\nu)(\vx) - (f_\nu \star f_\mu)(\vx),
\end{gather*}
and the Moyal product is def\/ined by \cite{nc}
\begin{gather*}
(f_\mu \star f_\nu)(\vx) \equiv e^{\frac i2 \theta_{ij}  \partial^i_x  \partial^j_y}f_\mu(\vx) f_\nu(\vy)|_{\vy \rightarrow \vx}.
\end{gather*}

In order to use some symplectic results, kinetic and potential parts of $\cal L$ need to be separated. This Lagrangian can be written as
\begin{gather*}
	{\cal L} = \frac 1{2m} \epsilon^{ij} f_i \dot f_j - V,
\end{gather*}
where $i,j = 1,2$, $\epsilon^{12}=1$ and
\begin{gather*}
	V = -\frac 12 f^\mu f_\mu + \frac 1m \ep^{ij}f_i\partial_jf_0 - \frac 3{4m} f_0 \ep^{ij}ie[f_i,f_j].
\end{gather*}

Now we will introduce two WZ f\/ields ($\theta$ and $\gamma$) and two unknown functions, def\/ining $\tilde \cl$:
\begin{gather*}
	\tilde {\cal L} (f_\mu, \dot f_\mu, \theta, \dot \theta, \gamma) = \frac 1{2m} \epsilon^{ij} f_i \dot f_j + (\Psi + \gamma)\dot \theta - \tilde V,
\end{gather*}
where
\begin{gather*}
	\Psi  \equiv  \Psi(f_\mu, \theta), \qquad
	\tilde V   \equiv   V + G + \frac 12 k \gamma \gamma, \qquad
	G  \equiv  G(f_\mu, \theta)
\end{gather*}
and $k$ is a constant. The dependence on the spatial derivatives is implicit in the equations above. The function $G$ satisf\/ies the condition $G(\theta =0)=0$.

The Lagrangian $\tilde \cl$ is not supposed to be explicitly invariant under some set of gauge transformations.
As it will be shown, a constraint must be added with this objective.

Let
\begin{gather*}
	(\tilde \xi^{ \alpha} )   =   \left(\begin{array}{cccc}f_0 & f^i & \theta & \gamma\end{array}\right), \qquad
	(\tilde a_\alpha)   =   \left(\begin{array}{cccc}0 & \frac 1{2m} \ep_{ij}f^i & \Psi + \gamma & 0\end{array}\right)
\end{gather*}
be the symplectic coordinates and momenta respectively, with $\alpha = 1,2,\dots ,5$. Thus, the symplectic matrix,
\begin{gather*}
	\tilde f_{\alpha \beta} (\vx,\vy) \equiv \frac{\delta \tilde a_\beta(\vy)}{\delta \tilde \xi^{\alpha}(\vx)} - \frac{\delta \tilde a_\alpha(\vx)}{\delta \tilde \xi^{\beta}(\vy)},
\end{gather*}
is given by
\begin{gather*}
\tilde f_{\alpha \beta} = \left(\begin{array}{cccc} 0 & 0 & \frac{\delta \Psi(\vy)}{\delta f_0(\vx)} & 0 \vspace{1mm}\\
0 & \frac {\ep_{ij}}m \delta (\vx - \vy) & \frac{\delta \Psi(\vy)}{\delta f^i(\vx)} & 0 \vspace{1mm}\\
- \frac{\delta \Psi(\vx)}{\delta f_0(\vy)} & -\frac{\delta \Psi(\vx)}{\delta f^j(\vy)} & \Theta_{xy} & -\delta (\vx -\vy) \vspace{1mm}\\
0 & 0 & \delta (\vx - \vy) & 0\end{array}\right),
\end{gather*}
where, as before,
\begin{gather*}
\Theta_{xy} \equiv \frac{\delta \Psi(\vy)}{\delta \theta(\vx)} - \frac{\delta \Psi(\vx)}{\delta \theta(\vy)}.
\end{gather*}

Let us choose the zero-mode as
\begin{gather*}
	(\tilde \nu^{\alpha}) = \left(\begin{array}{cccc} 1 & 0_{1 \times 2} & 0 & b\end{array}\right) ,
\end{gather*}
with $b$ constant. Except for the last two components, $(\tilde \nu^{\alpha})$ is the zero-mode of the symplectic matrix of $\cl$. The choice made in the last equation implies the following condition for $\Psi$,
\begin{gather}
	\label{psi0}
	\frac{\delta \Psi (\vy)}{\delta f_0(\vx)} = -b \delta(\vx - \vy) ,
\end{gather}
and with that zero-mode, we f\/ind the constraint
\begin{gather*}
\tilde \Omega(\vx)   \equiv   \int d^2y \, \tilde \nu^{\alpha} \frac {\delta \tilde V(\vy)}{\tilde \xi^{\alpha}(\vx)} \nonumber \\
	\phantom{\tilde \Omega(\vx)}{}   =   -f_0(\vx) + \frac 1m \ep^{ij} \partial_i f_j(\vx) - \frac 3{4m} \ep^{ij}ie[f_i,f_j](\vx) + \int d^2y \frac{\delta G(\vy)}{\delta f_0(\vx)} + bk\gamma (\vx).
\end{gather*}
Using $\Omega$ to express the constraint of the original theory and
\begin{gather*}
G_0(\vx) \equiv \int d^2y \frac{\delta G(\vy)}{\delta f_0(\vx)} ,
\end{gather*}
we can write
\begin{gather*}
	\tilde \Omega = \Omega + G_0 + bk \gamma.
\end{gather*}

Following the symplectic approach, let us insert this constraint into the kinetic part of the Lagrangian $\tilde \cl$. Hence, we have that,
\begin{gather}
	\label{l1}
	\tilde \cl^\1 = \frac 1{2m} \ep^{ij} f_i \dot f_j + (\Psi + \gamma) \dot \theta + \tilde \Omega \dot \lambda - \tilde V.
\end{gather}

With $\alpha = 1,2,\dots ,6$ and
\begin{gather*}
	(\tilde \xi^{\1 \alpha} )   =   \left(\begin{array}{ccccc}f_0 & f^i & \theta & \gamma & \lambda\end{array}\right), \qquad
	(\tilde a_\alpha^\1)   =   \left(\begin{array}{ccccc}0 & \frac 1{2m} \ep_{ij}f^i & \Psi + \gamma & 0 & \tilde \Omega\end{array}\right),
\end{gather*}
the symplectic matrix is
\begin{gather*}
\left(\tilde f^\1_{\alpha \beta}\right) =
\left(\begin{array}{ccccc} 0 & 0 & \frac{\delta \Psi(\vy)}{\delta f_0(\vx)} & 0 & \frac {\delta \tilde \Omega(\vy)}{\delta f_0(\vx)} \vspace{1mm}\\
0 & \frac {\ep_{ij}}m \delta (\vx - \vy) & \frac{\delta \Psi(\vy)}{\delta f^i(\vx)} & 0 & \frac {\delta \tilde \Omega(\vy)}{\delta f^i(\vx)} \vspace{1mm} \\
- \frac{\delta \Psi(\vx)}{\delta f_0(\vy)} & -\frac{\delta \Psi(\vx)}{\delta f^j(\vy)} & \Theta_{xy} & -\delta (\vx -\vy) & \frac {\delta G_0(\vy)}{\delta \theta (\vx)} \vspace{1mm} \\
0 & 0 & \delta (\vx - \vy) & 0 & kb\delta(\vx - \vy) \vspace{1mm} \\
- \frac {\delta \tilde \Omega(\vx)}{\delta f_0(\vy)} & -\frac {\delta \tilde \Omega(\vx)}{\delta f^j(\vy)} & -\frac {\delta G_0(\vx)}{\delta \theta (\vy)} & -kb \delta (\vx - \vy) & 0 \end{array}\right).
\end{gather*}

The zero-modes of $(\tilde f^\1_{\alpha \beta})$, which will be chosen, will turn out to be the gauge generators of the embedded theory. The structure of the selected zero-modes is
\begin{gather*}
	(\tilde \nu_\theta^{\alpha}(\vx))   =   \left(\begin{array}{ccccc}\rho_0 & \rho^i_x  & -kb & 0 & 1\end{array}\right), \qquad
	(\tilde \nu_\gamma^{ \alpha})   =   \left(\begin{array}{ccccc}1 & 0_{1 \times 2} & 0 & b & 0\end{array}\right).
\end{gather*}
As in the Proca model, our notation is such that the component $\rho_0$ is a constant and
$\rho_x^i \equiv a \partial_x^i$, where $a$ is another constant. So, at this point, we have not yet f\/ixed  the gauge generators altogether, i.e., $k$, $b$, $\rho_0$ and $\rho_x^i$ still have some freedom. A relation between $k$ and $b$ will be found but there is no other restriction.   The f\/inal answer will be quite general and, as a special case, we will f\/ind a St\"{u}ckelberg-like embedded theory.

We will assume the existence of a function $\Psi$ compatible with the zero-modes
$\tilde \nu_\theta$ and $\tilde \nu_\gamma$.
These objects need to be the gauge generators, so the function $G$ must agree with
\begin{gather*}
	\int d^2y \tilde \nu_\theta^{\alpha} \frac{\delta \tilde V(\vy)}{\delta \tilde \xi^{\1 \alpha}(\vx)} = 	\int d^2y \tilde \nu_\gamma^{ \alpha} \frac{\delta \tilde V(\vy)}{\delta \tilde \xi^{\1 \alpha}(\vx)} = 0.
\end{gather*}

For $\tilde \nu_\gamma$ there is no dif\/f\/iculty, its contraction with the gradient of $\tilde V$ is equal to $\tilde \Omega$, which is zero, accordingly to the kinetic part of $\tilde \cl^\1$. For $\tilde \nu_\theta$ we have the following dif\/ferential equation,
\begin{gather*}
	\int d^2y   \left \{ \rho_0 \left( \Omega(\vy) \delta(\vx -\vy) + \frac {\delta G(\vy)}{\delta f_0(\vx)} \right) + \rho_x^i \dirac \left( -f_i(\vy) + \frac 1m \ep_{ij} {\partial^j_y} f_0 (\vy)   \right. \right. \nonumber\\
\left. \left. \qquad{}- \frac {3ie}{2m} \ep_{ij}[f^j,f_0](\vy) \right) + \rho_x^i \frac{\delta G(\vy)}{\delta f^i(\vx)} - kb \frac{\delta G(\vy)}{\delta \theta(\vx)} \right \} = 0.
\end{gather*}

Writing
\begin{gather*}
	G = \sum_{n=1}^\infty {\cal G}_n
\end{gather*}
with ${\cal G}_n$ being proportional to $\theta^n$ or its spatial derivatives.  For the zeroth order in theta, we obtain
\begin{gather*}
	\int d^2y   \left \{ \rho_0 \Omega(\vy) \delta(\vx -\vy) + \rho_x^i \dirac \left( -f_i(\vy) + \frac 1m \ep_{ij}  {\partial^j_y} f_0 (\vy)   \right. \right. \nonumber \\
  \left. \left. \qquad{}- \frac {3ie}{2m} \ep_{ij}[f^j,f_0](\vy) \right) - kb \frac{\delta \cg_1(\vy)}{\delta \theta(\vx)} \right \} = 0,
\end{gather*}
whose solution is
\begin{gather*}
	\cg_1 = \frac \theta{kb} \left(-\rho^\mu f_\mu + \frac 1m \ep_{\mu i \lambda}   \rho^\mu \partial^i f^\lambda - \frac{3ie}{4m}\ep_{\mu \nu \lambda}   \rho^\mu[f^\nu,f^\lambda] \right).
\end{gather*}

For $\cg_2$ we have that,
\begin{gather*}
	\int d^2y \left \{ \rho^\mu_x \frac {\delta \cg_1(\vy)}{\delta f^\mu(\vx)} -kb \frac {\delta \cg_2(\vy)}{\delta \theta(\vx)} \right \}= 0,
\end{gather*}
hence
\begin{gather*}
	\cg_2 = -\frac 1{2k^2b^2} \left( \rho^\mu \theta \rho_\mu \theta + \frac {3ie}{2m} \ep_{\mu \nu \lambda}   f^\mu [\rho^\nu \theta, \rho^\lambda \theta] \right).
\end{gather*}

Finally, $\cg_3$ is given by
\begin{gather*}
	\int d^2y \left \{ \rho_x^\mu \frac {\delta \cg_2(\vy)}{\delta f^\mu(\vx)} -kb \frac {\delta \cg_3(\vy)}{\delta \theta(\vx)} \right \} = 0.
\end{gather*}
Although the above derivatives of $\cg_2$ do not vanish, when they are contracted with vector $(\rho^\mu)$ the result is zero, hence $\cg_3 = 0$. This result implies that
\begin{gather*}
	\cg_n = 0 \qquad \forall \, n \ge 3.
\end{gather*}

Now the function $G$ is known. Consequently $G_0$ can be determined as well,
\begin{gather*}
	G_0 = - \frac 1{kb} \left( \rho_0 \theta + \frac{3ie}{2m} \ep_{ij}   [\theta, \rho^i f^j] + \frac{3ie}{4kbm} \ep_{ij} [\rho^i \theta, \rho^j \theta] \right).
\end{gather*}

We have assumed that $\tilde \nu_\gamma$ and $\tilde \nu_\theta$ are zero-modes of matrix $\tilde f^\1$, now this condition will be used to f\/ind the function $\Psi$. Contracting $\tilde \nu_\theta$ with $\tilde f^\1$ and demanding this to be null, the following nontrivial equations emerge:
\begin{gather}
kb^2 = 1,\nonumber
\\
	\int d^2x  \left \{ \frac 1m \ep_{ij} \rho^i_x \dirac + \frac 1b \frac{\delta \Psi (\vx)}{\delta f^j (\vy)} - \frac 1m \ep_{ij}  {\partial^i_x} \dirac + \frac {3ie}{2m} \ep_{ij}[f^i(\vx),\dirac]   \right.\nonumber \\
\label{psii}
  \left. \qquad{} + \frac {3bie}{2m} \ep_{ij}[\theta (\vx), \rho_x^i \dirac] \right \} = 0
\end{gather}
and
\begin{gather}
	\label{psitheta}
	\int d^2x \left \{ \rho_x^i \frac{\delta \Psi (\vy)}{\delta f^i (\vx)} - \frac 1b \Theta_{xy} + \frac bm \ep_{ij} \left ( \rho_x^i {\partial^j_x} \dirac + \frac{3ie}{2} [\dirac, \rho^i_x f^j(\vx)] \right ) \right \} = 0,
\end{gather}
where the f\/irst equation have been used to achieve the last two. These, together with (\ref{psi0}), determines $\Psi$. The vector $\tilde \nu_\gamma$ does not generate any new condition to $\Psi$.

From equations (\ref{psi0}) and (\ref{psii}), except for a function dependent only on $\theta$, $\Psi$ can be computed. Equation (\ref{psitheta}), as it can be checked, is redundant. So, there is some arbitrariness left on $\Psi$, namely,  if $\Psi$ is a solution of above equations, the same is true for $\Psi(f_0,f^i,\theta) + f(\theta)$, where $f(\theta)$ is any function of $\theta$. However, this arbitrariness is not important, since such function only contributes to the surface term for the action~(\ref{l1}).

Therefore, within our purpose, $\Psi$ is
\begin{gather*}
	\Psi = \frac bm \ep_{ij} \left(\partial^i - \rho^i \right)f^j - \frac {3bie}{2m} \ep_{ij} \left (\frac 12 [f^i,f^j] + b[\theta, \rho^i f^j] \right ) - bf_0.
\end{gather*}

Hence, writing together all the results obtained before, the Lagrangian $\tilde \cl^\1$ is,
\begin{gather*}
	\tilde \cl^\1   = \frac 12 f^\mu f_\mu - \frac 1{4m} \ep^{\mu \nu \lambda} f_\mu F_{\nu \lambda} + \dot \lambda \tilde \Omega \nonumber\\
\phantom{\tilde \cl^\1   =}{}
+ \frac bm \ep_{ij} \left\{ \partial^i f^j - \frac {3ie}2 \left ( \frac 12 [f^i,f^j] + b[\theta, \rho^i f^j] \right ) - bf_0 + \gamma \right \}\dot \theta
  - \frac bm \ep_{i j} \rho_0 f^i \partial^j \theta \nonumber\\
\phantom{\tilde \cl^\1   =}{}
  + \frac {b^2}2 \rho^\mu \theta \rho_\mu \theta - b\theta \rho^\mu f_\mu + \frac {3ibe}{4m} \ep_{\mu \nu \lambda} \left( b f^\mu [\rho^\nu \theta, \rho^\lambda \theta] - \theta \rho^\mu [f^\nu,f^\lambda] \right) - \frac 1{2b^2} \gamma \gamma .   
\end{gather*}

Following the symplectic method \cite{mon}, the above Lagrangian has gauge symmetry with two free generators (both related with zero-modes $\tilde \nu_\theta$ and $\tilde \nu_\gamma$), which are
\begin{alignat}{3}
& \delta_{\vep_\theta}f_0 = \vep_\theta \rho_0 , \qquad &  & \delta_{\vep_\gamma} f_0 = \vep_\gamma, &  \nonumber\\
& \delta_{\vep_\theta}f^i = - \rho^i \vep_\theta, \qquad  &  & \delta_{\vep_\gamma} f^i = 0, &   \nonumber\\
& \delta_{\vep_\theta}\theta = - \vep_\theta \frac 1b,\qquad  & & \delta_{\vep_\gamma} \theta = 0, & \label{gg} \\
& \delta_{\vep_\theta}\gamma = 0,  \qquad & & \delta_{\vep_\gamma} \gamma = \vep_\gamma b, &   \nonumber\\
& \delta_{\vep_\theta}\lambda = \vep_\theta,\qquad  &  & \delta_{\vep_\gamma}\lambda = 0  ,&  \nonumber
\end{alignat}
the inf\/initesimal parameters are $\vep_\theta (\vx)$ and $\vep_\gamma (\vx)$. One can check that\footnote{Actually, $\delta_{\vep_\gamma} \tilde \cl^\1 = \tilde \Omega$, but we know from equations of motion that $\tilde \Omega = 0$.} $\delta_{\vep_\theta} \tilde \cl^\1 = \delta_{\vep_\gamma} \tilde \cl^\1 = 0$.

At this point we have already calculated the embedded version of the NC self-dual model. Our next step is to rewrite the Lagrangian and its gauge generators in another form.  The objective is to obtain a St\"{u}ckelberg-like Lagrangian. Hence, we are looking for a symmetry as
\begin{gather*}
	\delta_\vep f^\mu = \vep \partial^\mu, \qquad \delta_\vep \theta = - \frac 1b \vep.
\end{gather*}
Comparing this with (\ref{gg}), it is not hard to imagine that we have to eliminate $\gamma$ through $\gamma = b^2 \dot \theta$ and choose
\begin{gather*}
	(\rho^\mu) = \left(\begin{array}{cc} 0 & \partial^i\end{array}\right),
\end{gather*}
following the procedure explained in Section~\ref{sec:II}. Thus, the desired Lagrangian is found,
\begin{gather*}
	\cl_S =  \frac 12 \left(f_\mu - b\partial_\mu \theta\right) \left(f^\mu - b\partial^\mu \theta\right) \nonumber\\
 \phantom{\cl_S =}{} -  \frac 1{4m} \ep^{\mu \nu \lambda} \left(f_\mu - b \partial_\mu \theta \right) \left(\partial_\nu f_\lambda - \partial_\lambda f_\nu
 - ie[f_\nu -  b\partial_\nu \theta, f_\lambda - b \partial_\lambda \theta]\right) .
\end{gather*}

In this section we obtained a dual version, with gauge symmetry, of constrained and unconstrained f\/ield theory models, including the case of NC manifolds.
It is important to emphasize two remarkable features of the used method: it is easy to handle the NC part of the theory and the possibility of choosing, among the inf\/initesimal gauge generators, which gauge theory will be built in.
However, as said before we have to pay attention to the physical conditions governing the system.
Regarding the f\/irst point we can stress that other approaches \cite{ghosh} make use of the Seiberg--Witten map beforehand to set up a commutative version in order to handle the embedding. This may limit the range of applicability to certain powers of the parameter of the Moyal product.  The duality treated here however, is valid for   any power of the parameter of the Moyal product, since no restriction was necessary.

The commutative examples of this section considered constrained and unconstrained mo\-dels to illustrate the full power and generality of this technique.  Other embedding approaches are usually restricted to constrained models since they use the idea of constraint conversion to produce the gauge embedding. The embedding approach, on the other hand, only deals with the symplectic structure of the theory and does not depend on the previous existence of a~constrained structure to produce the gauge structure.  This f\/lexibility allowed us to investigate both commutative and NC theories indistinctly which brings great generality to the methodology.

\subsection[The massive NC $U(1)$ theory]{The massive NC $\boldsymbol{U(1)}$ theory}
\label{subsec:IIIA}

The Lagrangian density that governs the dynamics of NC massive $U(1)$ theory is
\begin{equation}
\label{equation1}
{\cal L} = - \frac{1}{4}F_{\mu\nu}F^{\mu\nu} + \frac{1}{2}m^2A^{\mu}A_{\mu},
\end{equation}
where the stress tensor in terms of the Moyal commutator is given by
\begin{equation}
\label{equation2}
F_{\mu\nu} = \partial_{\mu}A_{\nu} - \partial_{\nu}A_{\mu}
- ie[A_\mu,A_\nu],
\end{equation}
where
\begin{equation*}
[A_\mu,A_\nu] = A_{\mu} \star A_{\nu} - A_{\nu} \star A_{\mu},
\end{equation*}
and
\begin{gather*}
A_{\mu}(x) \star A_{\nu}(x) =\exp \left(\frac{i}{2}\theta^{\gamma\lambda}
\partial_{\gamma}^{x}\partial_{\lambda}^{y}\right)
A_{\mu}(x)A_{\nu}(y)\big\vert_{x=y},\nonumber \\
A_{\nu}(x) \star A_{\mu}(x) = \exp \left(\frac{i}{2}\theta^{\lambda\gamma}
\partial_{\lambda}^{x}\partial_{\gamma}^{y}\right)
A_{\nu}(x)A_{\mu}(y)\big\vert_{x=y},
\end{gather*}
where $\theta^{\gamma\lambda}$ is a real and antisymmetric constant matrix. In order to avoid causality and unitary problems in Moyal space, we take
$\theta^{0i} = 0$ \cite{Gomis}. Hence the $\star$ product of the gauge f\/ields into the stress tensor, given in equation~(\ref{equation2}), becomes
\begin{gather*}
A_{\mu}(x) \star A_{\nu}(x)  = \exp \left(\frac{i}{2}\theta^{ij}\partial_{i}^{x}\partial_{j}^{y}\right)
A_{\mu}(x)A_{\nu}(y)\big\vert_{x=y}, \nonumber\\
A_{\nu}(x) \star A_{\mu}(x)  = \exp\left(\frac{i}{2}\theta^{ji}\partial_{j}^{x}\partial_{i}^{y}\right)
A_{\nu}(x)A_{\mu}(y)\big\vert_{x=y}.
\end{gather*}

Now, we will reduce the second-order Lagrangian density (\ref{equation1}) into its f\/irst-order form, which is read as
\begin{gather*}
{\cal L}  =  \pi^i\dot A_i + A_0(\partial_i\pi^i + m^2A^0)
 + \frac{1}{2}\pi_i\pi^i
- ie\pi^i(A_0 \star A_i - A_i \star A_0)\nonumber\\
\phantom{{\cal L}  =}{} - \frac{1}{4}F_{ij}F^{ij} + \frac{1}{2}m^2A_iA^i
- \frac{1}{2}m^2A_0A^0,
\end{gather*}
where the canonical momentum $\pi_i$ is given by
\begin{gather*}
\pi_i  =  - F_{0i}
 =  -\dot A_i + \partial_iA_0 + ie(A_0 \star A_i - A_i \star A_0).
\end{gather*}

The symplectic f\/ields are $\xi^{(0)\alpha}=(A^i,\pi^i,A^0)$ and the zeroth-iterative symplectic matrix is
\begin{equation*}
f^{(0)} = \left(
\begin{array}{ccc}
0           & -\delta^{i}_{j} & 0 \\
\delta^{j}_{i}&         0     & 0 \\
0           &         0     & 0
\end{array}
\right) \delta({x} - {y}),
\end{equation*}
which is a singular matrix. It has a zero-mode that generates the following constraint
\begin{equation*}
\Omega(x) = \partial^{x}_{i}\pi^i(x) + m^2A^0(x) - ie\big[A_i(x),\pi^i(x)\big],
\end{equation*}
identif\/ied as being the Gauss law. Bringing back this constraint into the canonical part of the f\/irst-order Lagrangian density ${\cal L}^{(0)}$ using a Lagrangian multiplier ($\beta$), the f\/irst-iterated Lagrangian density, written in terms of $\xi^{(1)\alpha} = (A^i,\pi^i,A^0, \beta)$ is obtained as
\begin{gather*}
{\cal L}^{(1)}  =  \pi^{i}\dot A_i + \dot{\beta}\Omega + \frac{1}{2}\pi_i\pi^i - \frac{1}{4}F_{ij}F^{ij}
 +  \frac{1}{2}m^2A_iA^i - \frac{1}{2}m^2A_0A^0,
\end{gather*}
with the following symplectic f\/ields $\xi^{(1)\alpha}=(A^i,\pi^i,A^0,\beta)$.
Following the steps described in the sections before, the f\/irst-iterated symplectic matrix is obtained as,
\begin{equation*}
f^{(1)}=\left(
\begin{array}{cccc}
0           & -\delta^i_{j}\delta(x - y)  &  0   &   ie [\pi^i(y),\delta(x - y) ]\\
\delta^j_{i}\delta(x - y)  &         0     &   0    &   f_{\pi_i\beta}  \\
0         &         0     &   0    &     m^2\delta(x - y)   \\
ie[\delta(x - y),\pi^i(x)]        &        f_{\beta\pi_i}    &  -m^2 \delta(x - y)     &     0
\end{array}
\right),
\end{equation*}
where
\begin{gather*}
f_{\pi_i\beta}(x,y) = \partial^y_i\delta(x - y) + ie\big[\delta(x - y),A^i(y)\big].
\end{gather*}

This matrix is nonsingular and, as settle by the symplectic formalism, the Dirac brackets between the phase space f\/ields are acquired from the inverse of the symplectic matrix, namely,
\begin{gather*}
\lbrace A_i(x),A^j(y)\rbrace^*  =  0,\nonumber\\
\lbrace A_i(x),\pi^j(y)\rbrace^*  =  \delta_i^j\delta(x - y),\nonumber\\
\lbrace A_i(x),A_0(y)\rbrace^*  =  - \frac{1}{m^2}\left(\partial^x_i\delta(x - y) + \frac{ie}{m^2}[\delta(x - y),A_i(y)]\right),\nonumber\\
\lbrace \pi^i(x),A_0(y)\rbrace^*  =  \frac{ie}{m^2}[\delta(x - y),\pi^i(y)].
\end{gather*}

After all, we are ready to compute the symplectic embedding formalism of the theory.
Let us explain once more that the symplectic embedding process begins enlarging the phase space with the introduction of two WZ f\/ields $\gamma = \left(\begin{array}{cc}\eta&\pi_\eta\end{array}\right)$. Due to this, the original Lagrangian density~(\ref{equation1}) becomes
\begin{equation*}
{\tilde{\cal L}} = {\cal L} + {\cal L}_{\rm WZ},
\end{equation*}
where ${\cal L}_{\rm WZ}$ is a WZ counter-term which eliminates the noncommutativity of the theory. In agreement with the  symplectic embedding formalism, this new Lagrangian density must be reduced to its f\/irst-order form, namely,
\begin{gather*}
{\tilde{\cal L}}^{(0)}  =  \pi^i\dot A_i + \pi_\eta \dot \eta - {\tilde V}^{(0)},
\end{gather*}
where ${\tilde V}^{(0)}$ is the symplectic potential , given by
\begin{gather*}
{\tilde V}^{(0)} = - A_0(\partial_i\pi^i + m^2A^0)
- \frac{1}{2}\pi_i\pi^i
+ ie\pi^i(A_0 \star A_i + A_i \star A_0)
 \nonumber\\
 \phantom{{\tilde V}^{(0)} =}
+ \frac{1}{4}F_{ij}F^{ij} - \frac{1}{2}m^2A_iA^i + \frac{1}{2}m^2A_0A^0 + G(A^i,\pi^i,A^0,\gamma),
\end{gather*}
where $G\equiv G(A^i,\pi^i,A^0,\gamma)$ is an arbitrary function and is written as an expansion in terms of the WZ f\/ields as
\begin{equation*}
G(A^i,\pi^i,A^0,\gamma) = \sum_{n=0}^\infty {\cal G}^{(n)}(A^i,\pi^i,A^0,\gamma)\qquad \text{with}\quad  {\cal G}^{(n)}(A^i,\pi^i,A^0,\gamma)\sim (\gamma)^{n}.
\end{equation*}

The new symplectic variables are now given by ${\tilde\nu}^{(0)\alpha}=(A^i,\pi^i,A^0,\gamma)$ and the respective symplectic tensor is
\begin{gather*}
{\tilde f}^{(0)} = \left(\begin{array}{ccccc} 0 & - \delta_j^i\delta(x - y) & 0 & 0 & 0 \\
\delta_i^j\delta(x - y) &  0 & 0 & 0 & 0 \\
0 & 0 & 0 & 0 & 0 \\
0 & 0 & 0 & 0 & - \delta (x - y) \\
0 & 0 & 0 & \delta (x - y) & 0 \end{array}\right).
\end{gather*}
This singular matrix has a zero-mode, which is settle as
\begin{equation*}
{\tilde\nu}^{(0)} = \left(\begin{array}{ccccc}0 & 0 & 1 & 0 & 0 \end{array}\right).
\end{equation*}
This zero-mode when contracted with the symplectic potential generates the following constraint,
\begin{equation*}
\Omega (x) = \partial^x_i\pi^i(x) + m^2A^0(x) - i e[A^i(x),\pi_i(x)] + \int dy \frac{\delta G(y)}{\delta A^0(x)}.
\end{equation*}

In accordance with the symplectic formalism, this constraint must be introduced into the zeroth-iterative f\/irst-order Lagrangian density through a Lagrange multiplier $\zeta$, generating the next one,
\begin{equation*}
{\tilde{\cal L}}^{(1)} = \pi^i\dot A_i + \pi\dot \eta + \Omega\dot\zeta - {\tilde V}^{(1)},
\end{equation*}
with ${\tilde V}^{(1)}={\tilde V}^{(0)}\mid_{\Omega=0}$.  The symplectic vector is ${\tilde\xi}^{(1)} = (A^i,\pi^i,A^0,\zeta,\gamma)$ with the corresponding tensor given by
\begin{gather*}
{\tilde f}^{(1)} = \left(\begin{array}{cccccc}
0 & - \delta_i^j\delta(x - y) & 0 & \frac{\delta\Omega(y)}{\delta A^i(x)} & 0 & 0\vspace{1mm} \\
\delta_j^i\delta(x - y) &  0 & 0 & \frac{\delta\Omega(y)}{\delta \pi^i(x)} & 0 & 0 \vspace{1mm}\\
0 & 0 & 0 & \frac{\delta\Omega(y)}{\delta A^0(x)} & 0 & 0 \vspace{1mm}\\
- \frac{\delta\Omega(x)}{\delta A^j(y)} & - \frac{\delta\Omega(x)}{\delta \pi^j(y)} & - \frac{\delta\Omega(x)}{\delta A^0(y)} & 0 & - \frac{\delta\Omega(x)}{\delta \eta(y)} & - \frac{\delta\Omega(x)}{\delta \pi_\eta(y)} \vspace{1mm}\\
0 & 0 & 0 &   \frac{\delta\Omega(y)}{\delta \eta(x)} & 0 & - \delta(x - y) \vspace{1mm}\\
0 & 0 & 0 &  \frac{\delta\Omega(y)}{\delta \pi_\eta(x)}  & \delta(x - y) & 0\end{array}\right).
\end{gather*}

Hence, we are ready to remove the NC character of the original theory. To this end, it is necessary to assume that the symplectic matrix above is singular. Consequently, this matrix has its correspondent zero-mode, which is degenerated due to  the arbitrariness present in the matrix, which resides in the degenerated function $G$. This is not bad at all since it gives room to settle several approaches to eliminate the NC structure. Consequently, this opens up the possibility to obtain several commutative embedded representations for the NC model, which are all dynamically equivalent.
We believe that this represents an advantage of the symplectic embedding formalism. Therefore, we chose a convenient zero-mode as,
\begin{equation*}
{\tilde \nu}^{(1)} = \left(\begin{array}{cccccc}\partial^{x,i} & 0 & 0 & -1 & 1 & 1 \end{array}\right).
\end{equation*}

Using this zero-mode with the symplectic matrix above, as we did before, we can write that,
\begin{equation*}
\int d^3 x\,\, \nu^{(1) \tilde \alpha}(x) {\tilde f}^{(1)}_{\tilde \alpha \tilde \beta}(x,y) = 0,
\end{equation*}
and hence, we obtain the boundary condition ${\cal G}^{(0)}$ as
\begin{gather*}
{\cal G}^{(0)}(x) = - \frac 12 m^2 A_0(x)A^0(x) + ie\big[A^i(x),\pi_i(x)\big] A^0(x),
\end{gather*}
and some of the correction terms belong to ${\cal G}^{(1)},$ namely, $\pi_\eta A^0 - \eta A^0$. We note that the correction term ${\cal G}^{(n)}$ for $n\geq 2$ has no dependence on the temporal component of the potential f\/ield $A^0$. Thus, ${\cal G}^{(n)}\equiv{\cal G}^{(n)}(A^i,\pi^i,\gamma)$ for $n\geq 2$. This completes the f\/irst step of the symplectic embedding formalism.

Following the embedding procedure, after introducing these correction terms into the symplectic potential ${\tilde V}^{(1)}$, we will begin with the second step in order to reformulate the theory as a~gauge theory.
In the symplectic embedding formalism, the zero-mode ${\tilde \nu}^{(1)}$ does not produce a~constraint when contracted with the gradient of the symplectic potential, namely,
\begin{gather*}
\int d^3x \, \tilde\nu^{(1)\tilde\alpha}(y) \frac{\delta {\tilde V}^{(1)}(x)}{\delta {\tilde\xi}^{\tilde\alpha}(y)} = 0,
\end{gather*}
instead, it produces a general equation that allows the computation of the correction terms in powers of $\gamma$ enclosed into $G(A_i,\pi_i,A_0,\gamma)$. To compute the others linear correction terms in~$\gamma$,~${\cal G}^{(1)}$, we use the following relation (see~\eqref{2090})
\begin{gather*}
0  =  \int  d^3x  \,  \Bigg\{{-}\, ie\big[F_{ij}(x), A^i(x)\big]\partial^{j}_y\delta(x-y)
-   m^2\,A_j(x)\partial^{j}_y\delta(x-y) \nonumber \\
\hphantom{0=\int  d^3x  \, }{} +  ie\big[\pi_i(x), A_0(x)\big]\partial^{i}_y\delta(x-y)
 +  \frac{\delta{\cal G}^{(1)}(x)}{\delta\eta(y)} + \frac{\delta{\cal G}^{(1)}(x)}{\delta\pi_\eta(y)}\Bigg\} .
\end{gather*}
After a straightforward calculation, the complete linear correction term $\gamma$ is given by
\begin{gather*}
{\cal G}^{(1)}(x)   =  \pi_\eta(x) A^0(x) - \eta(x) A^0(x) + \big\{ie\partial^{j}_x\big[F_{ij}(x),A^i(x)\big]
 \nonumber \\
\phantom{{\cal G}^{(1)}(x)   =} {}+  ie\partial^{j}_x \big[\pi_j(x), A_0(x)\big]  +  m^2\partial^{j}_xA_j(x)\big\}\frac 12 (\eta(x) + \pi_\eta(x)) .
\end{gather*}

In order to compute the quadratic correction term, we use the following relation (see (\ref{2095}))
\begin{gather*}
\int d^3x  \left[\partial^{j}_x \left(\frac {\delta{\cal G}^{(1)}}{\delta A^j(x)} \right) + \frac{\delta{\cal G}^{(2)}(y)}{\delta\eta(x)} + \frac{\delta{\cal G}^{(2)}(y)}{\delta\pi_\eta(x)}\right] = 0,
\end{gather*}
and after a direct calculation, we have that
\begin{gather*}
{\cal G}^{(2)}(x)  = - \frac{ie}{4} F_{i,j}\big[\partial^{i}_x\eta(x), \partial^{j}_x\eta(x)\big]
 -  \frac 14 m^2 \partial^j_{x}\eta(x)\partial_j^{x}\eta(x) \nonumber \\
\phantom{{\cal G}^{(2)}(x)  =}{} -  \frac{e^2}{4} \big[\partial^{i}_x\eta(x), A_j(x)\big] \big[A_i(x), \partial^{j}_x\eta(x)\big]
 - \frac{e^2}{4}\big[A_i(x), \partial_j^{x}\eta(x)\big]
\big[A^i(x), \partial^{j}_x\eta(x)\big]\nonumber\\
\phantom{{\cal G}^{(2)}(x)  =}{} -  \frac{ie}{4} F_{i,j}\big[\partial^{i}_x\pi_\eta(x), \partial^{j}_x\pi_\eta(x)\big] - \frac 14 m^2 \partial^j_{x}\pi_\eta(x)\partial_j^{x}\pi_\eta(x)\\  
\phantom{{\cal G}^{(2)}(x)  =}{} - \frac{e^2}{4} \big[\partial^{i}_x\pi_\eta(x),\,A_j(x)\big] \big[A_i(x), \partial^{j}_x\pi_\eta(x)\big]
 - \frac{e^2}{4}\big[A_i(x), \partial_j^{x}\pi_\eta(x)\big]
\big[A^i(x), \partial^{j}_x\pi_\eta(x)\big] .\nonumber
\end{gather*}

The remaining corrections ${\cal G}^{(n)}$ for $n\geq 3$ are computed in analogous way (see (\ref{2100})) and we just write them down as
\begin{gather*}
{\cal G}^{(3)}(x)  =  \frac{e^2}{2} \big[A_i(x), \partial_j^{x}\eta(x)\big]
\big[\partial^{j}_x\eta(x), \partial^{y,i}\eta(x)\big]\nonumber\\
\hphantom{{\cal G}^{(3)}(x)  =}{}
+  \frac{e^2}{2}\big[A_i(x), \partial_j^{x}\pi_\eta(x)\big]
\big[\partial^{j}_x\pi_\eta(x), \partial^{y,i}\pi_\eta(x)\big] ,\nonumber \\
{\cal G}^{(4)}(x)  =  \frac{e^2}{8} \big[\partial_i^{x}\eta(x), \partial_j^{x}\eta(x)\big]
\big[\partial^{j}_x\eta(x),\,\partial^{i}_x\eta(x)\big]\nonumber\\
 \hphantom{{\cal G}^{(4)}(x)  =}{}
 +  \frac{e^2}{8} \big[\partial_i^{x}\pi_\eta(x), \partial_j^{x}\pi_\eta(x)\big]
\big[\partial^{j}_x\pi_\eta(x), \partial^{i}_x\pi_\eta(x)\big].  
\end{gather*}

Note that the fourth-order correction term has the WZ f\/ield dependence only, thus all correction terms ${\cal G}^{(n)}$ for $n\geq 5$ are zero. Then, the gauge invariant f\/irst-order Lagrangian density is given by
\begin{equation*}
{\tilde{\cal L}} = \pi^i(x)\dot A_i(x) + \pi(x)\dot \eta(x) - {\tilde {\cal H}},
\end{equation*}
where ${\tilde {\cal H}}$ is the gauge invariant Hamiltonian density, identif\/ied as being the symplectic poten\-tial~${\tilde V}^{(1)}$, namely,
\begin{gather*}
{\tilde {\cal H}}  =  {\tilde V}^{(1)} = - \frac{1}{2}\pi_i(x)\pi^i(x) + \frac{1}{4}F_{ij}F^{ij} - \frac{1}{2}m^2A_i(x)A^i(x) - ie\big[A^i(x), A_0(x)\big]\pi_i(x) \nonumber\\
\hphantom{{\tilde {\cal H}}  =  {\tilde V}^{(1)} =}{}
+ 2\pi_\eta(x) A^0(x) - 2\eta(x) A^0(x)
+ \frac{ie}{2}\partial^{j}_x\big[F_{ij}(x), A^i(x)\big]\eta(x) + \frac 12 m^2\partial^{j}_xA_j(x)\eta(x) \nonumber\\
\hphantom{{\tilde {\cal H}}  =  {\tilde V}^{(1)} =}{}
- \frac{ie}{2}\partial^{j}_x\big[\pi_i, A_0\big]\eta(x)
- \frac{ie}{4} F_{i,j}\big[\partial^{i}_x\eta(x), \partial^{j}_x\eta(x)\big]
- \frac 14 m^2 \partial^j_{x}\eta(x)\partial_j^{x}\eta(x)\nonumber \\
\hphantom{{\tilde {\cal H}}  =  {\tilde V}^{(1)} =}{}
- \frac{e^2}{4} \big[\partial^{i}_x\eta(x), A_j(x)\big] \big[A_i(x), \partial^{j}_x\eta(x)\big]
-\frac{e^2}{4}\big[A_i(x), \partial_j^{x}\eta(x)\big]
\big[A^i(x), \partial^{j}_x\eta(x)\big] \nonumber \\
\hphantom{{\tilde {\cal H}}  =  {\tilde V}^{(1)} =}{}
+ \frac{e^2}{2} \big[A_i(x), \partial_j^{x}\eta(x)\big]
\big[\partial^{j}_x\eta(x), \partial^{y,i}\eta(x)\big]
+ \frac{e^2}{8} \big[\partial_i^{x}\eta(x), \partial_j^{x}\eta(x)\big]
\big[\partial^{j}_x\eta(x), \partial^{i}_x\eta(x)\big]\nonumber\\
\hphantom{{\tilde {\cal H}}  =  {\tilde V}^{(1)} =}{}
+ \frac{ie}{2}\partial^{j}_x\big[F_{ij}(x), A^i(x)\big]\pi_\eta(x) + \frac 12 m^2\partial^{j}_xA_j(x)\pi_\eta(x)
- \frac{ie}{2}\partial^{j}_x\big[\pi_i, A_0\big]\pi_\eta(x)
\nonumber\\
\hphantom{{\tilde {\cal H}}  =  {\tilde V}^{(1)} =}{}
- \frac{ie}{4} F_{i,j}\big[\partial^{i}_x\pi_\eta(x), \partial^{j}_x\pi_\eta(x)\big]
- \frac 14 m^2 \partial^j_{x}\pi_\eta(x)\partial_j^{x}\pi_\eta(x)
\nonumber\\
\hphantom{{\tilde {\cal H}}  =  {\tilde V}^{(1)} =}{}
- \frac{e^2}{4} \big[\partial^{i}_x\pi_\eta(x), A_j(x)\big] \big[A_i(x) \partial^{j}_x\pi_\eta(x)\big]
-\frac{e^2}{4}\big[A_i(x), \partial_j^{x}\pi_\eta(x)\big]
\big[A^i(x), \partial^{j}_x\pi_\eta(x)\big]\nonumber \\
\hphantom{{\tilde {\cal H}}  =  {\tilde V}^{(1)} =}{}
+ \frac{e^2}{2}\big[A_i(x), \partial_j^{x}\pi_\eta(x)\big]
\big[\partial^{j}_x\pi_\eta(x), \partial^{y,i}\pi_\eta(x)\big] \nonumber\\
\hphantom{{\tilde {\cal H}}  =  {\tilde V}^{(1)} =}{}
+ \frac{e^2}{8} \big[\partial_i^{x}\pi_\eta(x), \partial_j^{x}\pi_\eta(x)\big]
\big[\partial^{j}_x\pi_\eta(x), \partial^{i}_x\pi_\eta(x)\big].
\end{gather*}

In order to complete the gauge invariant reformulation for the massive NC $U(1)$ theory, we compute the inf\/initesimal gauge transformations of the phase space coordinates. In agreement with the symplectic method, the zero-mode ${\tilde \nu}^{(1)}$ is the generator of the inf\/initesimal gauge transformations $(\delta {\cal O} = \varepsilon {\tilde \nu}^{(1)})$, which are given by
\begin{gather*}
\delta A_i  =  \partial^i \varepsilon,\qquad
\delta \pi^i  =  0,\qquad
\delta A_0  =  0,\qquad
\delta\eta  =  \varepsilon,\qquad
\delta\pi_\eta  =  \varepsilon,
\end{gather*}
where $\varepsilon(y)$ is an inf\/initesimal time-dependent parameter.
And with this result, we complete the Hamiltonian symplectic embedding of the massive NC $U(1)$ theory.

Summarizing, the  Hamiltonian density of the embedded version of the massive NC $U(1)$ theory was also obtained. A remarkable feature here is that the embedded version was obtained after a f\/inite number of steps of the iterative symplectic embedding process, which leads to a embedded Hamiltonian density with a f\/inite number of WZ terms.  It is important to regard that, by construction, these dif\/ferent embedding representations of the NC theory are dynamically equivalent, since the WZ gauge orbit is def\/ined by the zero-mode.

\section{Noncommutativity from the symplectic point of view}
\label{sec:IV}

In this section we will explain an extension of the symplectic embedding formalism that genera\-li\-zes the quantization by deformation introduced in \cite{DJEMAI1} in order to explore, with this new insight, how the NC geometry can be introduced into a commutative f\/ield theory.  To accomplish this, a~systematic way to introduce NC geometry into commutative systems, based on the symplectic approach and the Moyal product is presented.  However, this method describes precisely how to obtain a Lagrangian description for the NC version of the system.  To conf\/irm our approach, we use two well known systems, the chiral oscillator and some nondegenerate classical mechanics.  We will show precisely the NC contributions through this generalized symplectic method and obtain exactly the actions in the NC space found in the literature.

\subsection{Introducing noncommutativity through the generalized symplectic\\ formalism}
\label{subsec:IVA}

The quantization by deformation \cite{MOYAL1} consists in the substitution of the canonical quantization process by the algebra ${\cal A}_\hbar$ of quantum observables generated by the same classical one obeying the Moyal product, i.e., the canonical quantization
\begin{gather*}
\lbrace h, g\rbrace_{PB} = \frac{\partial h}{\partial \zeta_a} \omega_{ab} \frac{\partial g}{\partial \zeta_b}  \longrightarrow \frac {1}{\imath\hbar} [{\cal O}_h, {\cal O}_g] ,
\end{gather*}
with $\zeta=(q_i,p_i)$, is replaced by the $\hbar$-star deformation of ${\cal A}_0$, given by
\begin{gather*}
\lbrace h, g\rbrace_{\hbar} = h*_\hbar g - g*_\hbar h ,
\end{gather*}
where
\begin{gather*}
(h *_\hbar g)(\zeta) = \exp \left\{\frac{\imath}{2}\hbar \omega_{ab}\partial^a_{(\zeta_1)}\partial^b_{(\zeta_2)} \right\}h(\zeta_1)g(\zeta_2)|_{\zeta_1=\zeta_2=\zeta}
 ,
\end{gather*}
with $a,b=1,2,\dots,2N$ and with the following classical symplectic structure
\begin{gather*}
\omega_{ab} = \left(\begin{array}{ccc}0 & \delta_{ij}\cr -\delta_{ji} & 0\end{array}\right)\qquad \text{with}\quad i,j=1,2,\dots,N ,
\end{gather*}
that satisf\/ies the relation below
\begin{gather*}
\omega^{ab}\omega_{bc} = \delta^a_c .
\end{gather*}

The quantization by deformation can be generalized assuming a generic classical symplectic structure $\Sigma^{ab}$. In this way the internal law will be characterized by $\hbar$ and by another deformation parameter (or more). As a consequence, the $\Sigma$-star deformation of the algebra becomes{\samepage
\begin{gather*}
(h *_{\hbar\Sigma} g)(\zeta) =  \exp \left\{\frac{\imath}{2}\hbar \Sigma_{ab}\partial^a_{(\zeta_1)}\partial^b_{(\zeta_2)}\right\}h(\zeta_1)g(\zeta_2)|_{\zeta_1=\zeta_2=\zeta} ,
\end{gather*}
with $a,b=1,2,\dots,2N$.}

This new star-product generalizes the algebra among the symplectic variables  in the following way
\begin{gather*}
\lbrace h, g\rbrace_{\hbar\Sigma} = \imath\hbar\Sigma_{ab} .
\end{gather*}

In \cite{DJEMAI1,DJEMAI2}, the authors proposed a quantization process in order to transform the NC classical mechanics into the NC quantum mechanics, through the generalized Dirac quantization,
\begin{gather*}
\lbrace h, g\rbrace_{\Sigma} = \frac{\partial h}{\partial \zeta_a} \Sigma_{ab} \frac{\partial g}{\partial \zeta_b}  \longrightarrow \frac {1}{\imath\hbar} [{\cal O}_h, {\cal O}_g]_{\Sigma} .
\end{gather*}
The relation above can also be obtained through a particular transformation of the usual classical phase space, namely,
\begin{gather}
\label{00080}
\zeta^\prime_a = T_{ab} \zeta^b ,
\end{gather}
where the transformation matrix is
\begin{gather}
\label{00090}
T = \left(\begin{array}{cc} \delta_{ij} & - \frac 12 \theta_{ij} \\  \frac 12 \beta_{ij}  & \delta_{ij}\end{array}\right) ,
\end{gather}
where $\theta_{ij}$ and $\beta_{ij}$ are antisymmetric matrices. As a consequence, the original Hamiltonian becomes
\begin{gather*}
{\cal H}(\zeta_a) \longrightarrow {\cal H}(\zeta^\prime_a) ,
\end{gather*}
where the corresponding symplectic structure is
\begin{gather}
\label{00110}
\Sigma_{ab} = \left(\begin{array}{cc}\theta_{ij} & \delta_{ij}+\sigma_{ij} \\ -\delta_{ij}-\sigma_{ij} & \beta_{ij}\end{array}\right) ,
\end{gather}
with $\sigma_{ij} = - \frac18 [\theta_{ik}\beta_{kj} + \beta_{ik}\theta_{kj}]$. Due to this, the commutator relations look like
\begin{gather}
\label{00115}
\big[q^\prime_i, q^\prime_j\big]  =  \imath\hbar\theta_{ij} ,\qquad
\big[q^\prime_i, p^\prime_j\big]  =  \imath\hbar (\delta_{ij} + \sigma_{ij}) ,\qquad
\big[p^\prime_i, p^\prime_j\big]  =  \imath\hbar\beta_{ij} .
\end{gather}

We believe that at this point it is clear that here we are only analyzing systems where this symplectic algebra (\ref{00115}) involves only constants.  It is worthwhile to mention that there are systems where the symplectic algebra (\ref{00115}) involves phase space dependent quantities, rather than just constants.  For instance, we can mention the NC Landau problem (for example in \cite{horvathy} and references therein).
A particle in the NC plane, coupled to a constant magnetic f\/ield and an electric potential will possess an algebra similar to~(\ref{00115}).  A phase space dependent algebra occurs for a nonconstant magnetic f\/ield, as discussed in~\cite{horvathy}.   This problem can be object for future analysis.

Notice that a Lagrangian formulation was not given. Now, we propose a new systematic way to obtain a NC Lagrangian description for a commutative system. In order to achieve our objective, the symplectic structure $\Sigma_{ab}$ must be f\/ixed f\/irstly and subsequently, the inverse of $\Sigma_{ab}$ must be computed. As a consequence, an interesting problem arise: if there are some constant ({\it Casimir invariants}) in the system, the symplectic structure has a zero-mode, given by the gradient of these {\it Casimir invariants}. Hence, it is not possible to compute the inverse of~$\Sigma_{ab}$. However, in~\cite{CNWO} this kind of problem was solved. On the other hand, if~$\Sigma_{ab}$ is nonsingular, its inverse can be obtained solving the relation below
\begin{gather}
\label{00120}
\int \Sigma_{ab}(x,y) \Sigma^{bc}(y,z) d y = \delta_a^c\delta(x-z) ,
\end{gather}
which generates a set of dif\/ferential equations, since $\Sigma^{ab}$ is an unknown two-form symplectic tensor obtained from the following f\/irst-order Lagrangian
\begin{gather}
\label{00130}
{\cal L} = A_{\zeta^\prime_a} \dot\zeta^{\prime a} - V(\zeta^\prime_a) ,
\end{gather}
as being
\begin{gather}
\label{00140}
\Sigma^{ab}(x,y) = \frac {\delta A_{\zeta^\prime_b}(x)}{\delta \zeta^\prime_a(y)} - \frac {\delta A_{\zeta^\prime_a}(x)}{\delta \zeta^\prime_b(y)} .
\end{gather}
Due to this, the one-form symplectic tensor, $A_{\zeta^\prime_a}(x)$, can be computed and subsequently, the Lagrangian description, equation~(\ref{00130}), is obtained also. In order to compute $A_{\zeta^\prime_a}(x)$, equations~(\ref{00120}) and~(\ref{00140}) are used, which generates the following set of dif\/ferential equations
\begin{gather}
\theta_{ij} B_{jk}(x,y) + \left(\delta_{ij}+\sigma_{ij}\right)A_{jk}(x,y)  =  \delta_{ik}\delta(x-y) ,\nonumber\\
A_{jk}(x,y)\theta_{ji}  + \left(\delta_{ij}+\sigma_{ij}\right)C_{jk}(x,y)  =  0 ,\nonumber\\
- \left(\delta_{ij} + \sigma_{ij}\right)B_{jk}(x,y) + \beta_{ij}A_{jk}(x,y)  =  0 , \nonumber\\
A_{kj}(x,y)\left(\delta_{ji} + \sigma_{ji}\right) + \beta_{ij} C_{jk}(x,y)  =   \delta_{ik}\delta(x-y) ,\label{00150}
\end{gather}
where
\begin{gather}
B_{jk}(x,y)  =  \left(\frac {\delta A_{q^\prime_j}(x)}{\delta q^\prime_k(y)} - \frac {\delta A_{q^\prime_k}(x)}{\delta q^\prime_j(y)}\right) ,\qquad
A_{jk}(x,y)  =  \left(\frac {\delta A_{p^\prime_j}(x)}{\delta q^\prime_k(y)} - \frac {\delta A_{q^\prime_k}(x)}{\delta p^\prime_j(y)}\right) ,\nonumber\\
C_{jk}(x,y)  =  \left(\frac {\delta A_{p^\prime_j}(x)}{\delta p^\prime_k(y)} - \frac {\delta A_{p^\prime_k}(x)}{\delta p^\prime_j(y)}\right) .\label{00160}
\end{gather}
From the set of dif\/ferential equations (\ref{00150}), and the equations above (\ref{00160}), we compute the quantities $A_{\zeta^\prime_a}(x)$.

As a consequence, the f\/irst-order Lagrangian can be written as
\begin{gather}
\label{00185}
{\cal L} = A_{\zeta^\prime_a} \dot\zeta^\prime_a - V(\zeta^\prime_a) .
\end{gather}
Notice that, despite (\ref{00130}) and (\ref{00185}) have the same form, in~(\ref{00185}) the $A_{{\zeta'}_a}$ are completely computed through the solution of the system~(\ref{00150}).   In both we have a NC version of the theory as a consequence of the deformation in~(\ref{00090}) and its corresponding symplectic structure in~(\ref{00110}).   This will be clarif\/ied through the examples in the next section.

\subsection{Examples}
\label{subsec:IVB}

In this section  we will use the formalism developed above in two well known mechanical systems.  The f\/irst one is the chiral oscillator, which has a close relationship with the Floreanini--Jackiw version of the chiral boson \cite{FJa} through the mapping used in \cite{bazeia}.  The other one is the so-called nondegenerate mechanics \cite{alexei}.   We will show precisely that the results obtained with our formalism coincide with the ones depicted in  both systems, which conf\/irms the ef\/fectiveness of the method described in Section~\ref{sec:II}.

\subsubsection{The chiral oscillator}
\label{subsubsec:IVA}

Let us consider a two-dimensional model which has a reduced phase space. For this reason, the symplectic coordinates are given by $\zeta^\prime_a = (q^\prime_i)$, with $a=i=1,2$, and the canonical momenta conjugated to $q^\prime_i$ are not present. With this concept in mind, the NC algebra given in (\ref{00115}) is comprised only by the f\/irst element.  Therefore, following the procedure, the matrix $\Sigma_{ab}$ def\/ined in (\ref{00110}) now has only one element.  Then we consider the symplectic structure as being
\begin{gather*}
\Sigma_{ij}  =  \theta_{ij} =   \theta\,\epsilon_{ij} ,
\end{gather*}
where $\theta$ is the measure of the noncommutativity.  This reduces the set of dif\/ferential equations, given in equation~(\ref{00150}), to
\begin{gather}
\label{00200}
\frac {\delta A_{q^\prime_j}(x)}{\delta q^\prime_k(y)} - \frac {\delta A_{q^\prime_k}(x)}{\delta q^\prime_j(y)}  =  \theta_{ij}^{-1} = -  \theta\,\epsilon_{ij} .
\end{gather}
Notice that the prime is not the spatial derivative, it was def\/ined in~(\ref{00080}).

Now it is easy to see that the equation (\ref{00200}) has the following solution,
\begin{gather}
\label{00210}
A_{q^\prime_i} = -  \frac 12  \theta\,\epsilon_{ij} q^\prime_j .
\end{gather}
Substituting (\ref{00210}) in (\ref{00130}), the f\/irst-order Lagrangian is given by
\begin{gather*}
{\cal L} =  - \frac 12  \theta \epsilon_{ij} \dot q^\prime_i  q^\prime_j - V(q^\prime_j) .
\end{gather*}

We can assume that the symplectic potential is
\begin{gather*}
 V(q^\prime_j) = \frac{k{q'}_j^2}{2} .
\end{gather*}

 Thus, we have the mechanical version of the FJ chiral boson, namely, the chiral oscillator~(CO)~\cite{banerjee2}
\begin{gather}
\label{00240}
{\cal L} =  - \frac 12  \theta \epsilon_{ij} \dot q^\prime_i  q^\prime_j - \frac{k{q^\prime_j}^2}{2} .
\end{gather}
where dif\/ferent signs in $\theta$ will correspond to dif\/ferent chiralities, similarly as obtained in~\cite{banerjee2} (also studied in~\cite{banerjee3}).

To make an analogy of this model with a well known model for the chiral boson ($k=1$) let us make the following map using the relations described in~\cite{bazeia} given by,
\begin{gather*}
\partial_t \phi  \leftrightarrow  \partial_t {q'}_j ,\qquad
\partial_x \phi  \leftrightarrow   \theta\,\epsilon_{ij} {q'}_j ,
\end{gather*}
and with this map implemented in (\ref{00240}), it can be seen directly that the FJ chiral boson model~\cite{FJa} was obtained.

Although the chiral oscillator was discussed in details in various contexts (for instance, in~\cite{banerjee2,banerjee3} and references therein), the purpose of this specif\/ic example is only to illustrate our method of introducing the noncommutativity via the symplectic method.

\subsubsection{The nondegenerate mechanics}
\label{subsubsec:IVB}

With the understanding of the preceding example, it is now easy to see the procedure in a more complicated case, where the $\Sigma_{ab}$ matrix is bigger than before.

In \cite{alexei} it was introduced a NC version of an arbitrary nondegenerate mechanical system whose action can be written as
\begin{gather}
\label{00266}
S = \int dt\,L \big( q^A,\dot{q}^A\big) ,
\end{gather}
with the conf\/iguration space variables $q^A(t)$, $A=1,2,\dots,n$ and no constraints in the Hamiltonian formulation.

We consider now the following symplectic structure
\begin{gather*}
\Sigma_{\alpha\beta}  =  \left(\begin{array}{cc}- 2 \theta_{ij} & \delta_{ij} \\ - \delta_{ji} & 0\end{array}\right) ,
\end{gather*}
where, from (\ref{00110}), we can see that $\sigma_{ij}=\beta_{ij}=0$.  Using (\ref{00120}) we can construct the following matricial equation,
\begin{gather*}
\left(\begin{array}{cc}- 2 \theta_{il} & \delta_{il} \\ - \delta_{il} & 0\end{array}\right)
\left(\begin{array}{cc}\Sigma^{q_l\,q_j} & \Sigma^{q_l\,p_j} \\ \Sigma^{p_l\,q_j} & \Sigma^{p_l\,p_j}\end{array}\right) =
\left(\begin{array}{cc}\delta_i^{\;j} & 0 \\ 0 & \delta_i^{\;j}\end{array}\right)
\end{gather*}
and we can write that
\begin{gather}
- 2 \theta_{il} \Sigma^{q_l q_j}  +  \delta_{il} \Sigma^{p_l q_j}  =  \delta_i^{\;j} ,\qquad
\delta_{il} \Sigma^{q_l q_j}   =  0 ,\nonumber\\
- 2 \theta_{il} \Sigma^{q_l p_j} +  \delta_{il} \Sigma^{p_l p_j}  =  0 ,\qquad
-\delta_{il} \Sigma^{q_l p_j}  =  \delta_i^{\;j} .\label{00248}
\end{gather}

Solving (\ref{00248}) we have that
\begin{gather*}
\Sigma^{q_i q_j} = 0 , \qquad
\Sigma^{p_i q_j} = \delta_{ij} ,\qquad
\Sigma^{p_i p_j} = - 2 \theta_{ij} .
\end{gather*}
Hence
\begin{gather*}
\frac {\delta A_{q_j}(x)}{\delta q_i(y)}  -  \frac {\delta A_{q_i}(x)}{\delta q_j(y)}  = 0 ,\qquad
\frac {\delta A_{q_j}(x)}{\delta p_i(y)}  -  \frac {\delta A_{p_i}(x)}{\delta q_j(y)}  =  \delta_{ij} ,\nonumber\\
\frac {\delta A_{p_j}(x)}{\delta p_i(y)}  -  \frac {\delta A_{p_i}(x)}{\delta p_j(y)}  =  - 2 \theta_{ij} .
\end{gather*}

A convenient solution of this system is
\begin{gather*}
A_{q_i}  =  {1\over2} p_i + {1\over2} q_i ,\qquad
A_{p_i}  =  \theta_{im} p_m - {1\over2} q_i .
\end{gather*}

Finally, we can construct our f\/irst-order Lagrangian as
\begin{gather}
L  =  \left( {1\over2} p_i + {1\over2} q_i \right) \dot{q}_i +\left( \theta_{im} p_m - {1\over2} q_i \right) \dot{p}_i - V (q,\dot{q})  =  p_i \dot{q}_i + \dot{p}_i \theta_{ij} p_j - V' (q,\dot{q}) ,\label{00160a}
\end{gather}
where $V' (q,\dot{q})=V(q,\dot{q}) + {1\over2}q_i \dot{q}_i$ and $L$ is the same Lagrangian obtained in \cite{alexei} (in equation~(4) in~\cite{alexei}, the $H_0(q^A,v_A)$ is our $V(q,\dot{q})$). In few words we can say that the quantization of this system takes us to quantum mechanics with the ordinary product substituted by the Moyal product, similarly to the case of a particle on a NC plane~\cite{alexei}.

The Lagrangian (\ref{00160a}) is the NC version of the nondegenerate mechanical system described by the Lagrangian $L=L(q_i,\dot{q}_i)$~\cite{alexei}.   It is easy to see that (\ref{00160a}) has the same number of physical degrees of freedom as the initial system $S$, equation (\ref{00266}).  It can be demonstrated also that the equations of motion of the NC system are the same as for the initial system $S$, modulo the term which is proportional to the parameter $\theta^{AB}$.  Finally, we can say that the conf\/iguration space variables have the NC brackets: $\{q^A,q^B\} = -2 \theta^{AB} $ \cite{alexei}.

To end this section and consequently this work we will make some considerations
To deform a system by substituting the classical product by the Moyal product comprises essentially the usual embedding of a commutative system in a NC conf\/iguration space.  The f\/inal system is now recognized as a NC theory.  The last one has been investigated intensively in the literature.

In order to improve the knowledge of non-perturbative processes on how to obtain ef\/fectively a NC theory, the authors in \cite{DJEMAI1}  discuss the passage from classical mechanics to quantum mechanics and then to NC quantum mechanics, which allows one to obtain the associated NC classical mechanics.

We believe that with the symplectic formalism we can give a step further. It was proposed an alternative new way to obtain NC models, based on the symplectic approach. An interesting feature on this formalism lies on the symplectic structure, which is def\/ined at the beginning of the process. The choice of the symplectic structure, subsequently, def\/ines the NC geometry of the model and the Planck's constant enters the theory via Moyal product. This formalism also describes precisely how to obtain a Lagrangian description for the NC version of the system.

To illustrate the method, we used a chiral oscillator \cite{banerjee2,banerjee3} in the NC phase space that is equivalent to the Floreanini--Jackiw chiral boson through a convenient mapping.

We talked also about the NC version of an arbitrary nondegenerate mechanical system which has no constraints in the Hamiltonian formulation and where now, the conf\/iguration space variables have the NC brackets $\{q^A,q^B\}=-2 \theta^{AB}$. The result coincides with the ones in the literature.

It is important to stress that the procedure deals only with non-constrained systems.   A possible work of research is the investigation of how NC geometry can be introduced into constrained systems via symplectic approach. We believe that the method can bring new insights into this issue also.

\appendix

\section{Some properties of the Moyal product}

In this appendix we list some properties that we use in this paper.
\begin{gather*}
\int d^nx\, \phi_1 \star \phi_2 = \int d^nx\, \phi_1\phi_2 = \int d^nx\, \phi_2 \star \phi_1,\\
(\phi_1\star\phi_2)\star\phi_3 = \phi_1\star(\phi_2\star\phi_3) = \phi_1\star\phi_2\star\phi_3,    \\
\int d^nx\, \phi_1\star\phi_2\star\phi_3 = \int d^nx\, \phi_2\star\phi_3\star\phi_1
= \int d^nx\, \phi_3\star\phi_1\star\phi_2,   \\
\int d^nx\,  [ [A,B ],C ]\star D = \int d^nx\,  [A,B ]\star  [C,D ]
= \int d^nx\, [A,B ]  [C,D ].
\end{gather*}

\subsection*{Acknowledgments}

EMCA would like to thank the hospitality and kindness of the Dept. of Physics of the Federal University of Juiz de Fora where part of this work was done.
This work was supported in part by Funda\c c\~ao de Amparo a Pesquisa do Estado de Minas Gerais (FAPEMIG) and Conselho Nacional de Desenvolvimento Cient\'\i f\/ico e Tecnol\'ogico (CNPq), Brazilian Research Agencies.


\pdfbookmark[1]{References}{ref}
\LastPageEnding

\end{document}